\documentclass{nature2}

\pdfoutput=1

\usepackage{verbatim}

\usepackage{pdflscape} 
\usepackage{amssymb}
\usepackage{aas_macros}
\usepackage[T1]{fontenc}
\usepackage{graphicx}        
\usepackage[utf8]{inputenc}
\usepackage{hyperref}
\usepackage{bookmark}
\usepackage{bm}
\usepackage{amsfonts}
\usepackage{amsmath}

\newcommand{\figu}{Figure~}

\newcommand{\eq}{Equation~}
\newcommand{\eqs}{Equations~}
\newcommand{\sect}{Section~}

\usepackage{xpatch}

\newcounter{mybib}

\newcommand{\ergse}{{\rm erg\, s^{-1}}}

\newcommand{\sis}{$\sigma$}
\newcommand{\sise}{\sigma}

\newcommand{\mbh}{$M_{\rm bh}$}
\newcommand{\mbhe}{M_{\rm bh}}

\newcommand{\mh}{$M_{\rm halo}$}
\newcommand{\mhe}{M_{\rm halo}}
\newcommand{\mhalo}{$M_{\rm halo}$}

\newcommand{\mstar}{$M_{\rm star}$}
\newcommand{\mstare}{M_{\rm star}}

\newcommand{\msune}{M_{\odot}}

\newcommand{\fobse}{f_{\rm obsc}}

\newcommand{\kbe}{k_{\rm bol}}

\newcommand{\ndash}{-}

\newcommand{\epsie}{\varepsilon}
\newcommand{\epsi}{$\varepsilon$}

\title{Constraining black hole-galaxy scaling relations from the large-scale clustering of Active Galactic Nuclei and implied
mean radiative efficiency}

\author{Francesco Shankar$^{1}$, Viola Allevato$^{2,3,4}$, Mariangela Bernardi$^{5}$, Christopher Marsden$^{1}$, Andrea Lapi$^{6,7}$, Nicola Menci$^{8}$, Philip J. Grylls$^{1}$, Mirko Krumpe$^{9}$, Lorenzo Zanisi$^{1}$, Federica Ricci$^{10}$, Fabio La Franca$^{11}$, Ranieri D. Baldi$^1$, Jorge Moreno$^{12,13}$, Ravi K. Sheth$^{5}$}

\begin{document}

\maketitle

\begin{affiliations}
 \item Department of Physics and Astronomy, University of Southampton, Highfield, SO17 1BJ, UK
 \item Scuola Normale Superiore, Piazza dei Cavalieri 7, I-56126 Pisa, Italy
 \item Department of Physics and Helsinki Institute of Physics, Gustaf H\"allstr\"omin katu 2a, 00014 University of Helsinki, Finland
 \item INAF-Osservatorio Astronomico di Bologna, Via Ranzani 1, 40127 Bologna, Italy
 \item Department of Physics and Astronomy, University of Pennsylvania, 209 South 33rd St, Philadelphia, PA 19104, USA
 \item SISSA, Via Bonomea 265, I-34136 Trieste, Italy
 \item IFPU - Institute for fundamental physics of the Universe, Via Beirut 2, 34014 Trieste, Italy
 \item INAF–Osservatorio Astronomico di Roma, via Frascati 33, 00078 Monteporzio Catone, Italy
 \item Leibniz-Institut f\"{u}r Astrophysik Potsdam (AIP), An der Sternwarte 16, D-14482 Potsdam, Germany
 \item Instituto de Astrof\'{\i}sica and Centro de Astroingenier\'{\i}a, Facultad de Física, Pontificia Universidad Cat\'{o}lica de Chile, Casilla 306, Santiago 22, Chile
 \item Dipartimento di Matematica e Fisica, Universit\`{a} Roma Tre, via della Vasca Navale 84, I-00146 Roma, Italy
 \item Department of Physics and Astronomy, Pomona College, Claremont, CA 91711, USA
 \item Harvard-Smithsonian Center for Astrophysics, 60 Garden Street, Cambridge, MA, 02138, USA
\end{affiliations}

\begin{abstract}
A supermassive black hole has been found at the centre of nearly every galaxy observed with sufficient sensitivity. The masses of these black holes are observed to increase with either the total mass or the mean (random) velocity of the stars in their host galaxies. The origin of these correlations remains elusive.
Observational systematics and biases severely limit our knowledge of the local demography of supermassive black holes thus preventing accurate model comparisons and progress in this field. Here we show that the large-scale spatial distribution of local active galactic nuclei (AGN), believed to be accreting supermassive black holes, can constrain the shape and normalization of the black hole-stellar mass relation thus bypassing resolution-related observational biases. In turn,
our results can set more stringent constraints on the so-called ``radiative efficiency'', $\epsie$, a fundamental parameter describing the inner physics of supermassive black holes that is closely linked to their spin, geometry, and ability to release energy. The mean value of $\epsie$ can be estimated by comparing the average total luminous output of AGN with the relic mass density locked up in quiescent supermassive black holes at galaxy centres today. For currently accepted values of the AGN obscured fractions and bolometric corrections, our newest estimates of the local supermassive black hole mass density favour mean radiative efficiencies of $\epsie \sim 10-20\%$, suggesting that the vast majority of supermassive black holes are spinning moderately to rapidly. With large-scale AGN surveys coming online, our novel methodology will enable even tighter constraints on the fundamental parameters that regulate the growth of supermassive black holes.
\end{abstract}

In the traditional picture, quiescent supermassive black holes at the centres of local galaxies today are the relics of a previous phase of active gas accretion in which they shone as Active Galactic Nuclei (AGN)\cite{Rees84}. The brightness of an AGN depends on how efficiently the gas orbiting the black hole can radiate.  In principle, the mean radiative efficiency $\epsie$ (and hence black hole spin\cite{Bardeen72}) can be constrained by equating\cite{Soltan} the mass accreted by all AGN over all times with the mass density in supermassive black holes today\cite{Salucci99,Marconi04,Shankar13acc,Aversa15}. The accreted mass scales roughly with the inverse of the mean radiative efficiency, because a lower radiative efficiency requires a higher mass accretion rate to produce the observed AGN luminosity.  In this scenario, the same total AGN emissivity can be matched either by a small black hole mass density today but a large radiative efficiency, or vice-versa.  This has fueled interest in estimating the local black hole mass density or ``mass function'', i.e. the number density per unit comoving volume and bin of black hole mass.

Quiescent black holes are difficult to observe, so large samples are not available.  Therefore, the black hole mass function is usually estimated in two steps.  First, one uses the small sample that is available to calibrate how a black hole's mass correlates with other properties of its host galaxy, such as its stellar mass \mstar\ or the velocity dispersion of its stars \sis.  Since \mstar\ and \sis\ can be estimated in much larger samples, one uses these ``scaling relations'' to transform the comoving number density of \mstar\ or \sis\ into that for black hole mass\cite{Salucci99}.

Unfortunately, despite its elegance, the method described above suffers from systematic uncertainties that prevent a robust estimate of \epsi, the mean radiative efficiency of supermassive black holes. The main difficulty is to estimate the local scaling relations reliably\cite{Shankar16BH}. Over the past decade, a number of groups have put forward a variety of rather different relations. For instance, there are several claims that black holes hosted in galaxies belonging to different morphological classes follow different scaling relations\cite{KormendyHo,Davis18}. In addition, active and quiescent black holes in the local Universe appear to follow somewhat differently normalized scaling relations, with the degree of the offset depending on the type of galaxies, AGN sample, or scaling relation considered\cite{busch2014low,ReinesVolonteri15,Shankar19BH}.  Furthermore, limitations on instrumental capabilities or other observational effects may systematically bias the observed scaling relations away from the intrinsic ones.  This is a concern especially in quiescent early-type galaxies with dynamically-measured black hole masses\cite{Bernardi07,DaiMbhSigma,Shankar16BH}. In fact, recent work\cite{Shankar19BH} suggests that such a bias may explain much of the reported difference between active and quiescent scaling relations:  the intrinsic relations may be much more similar than they appear.

In what follows, we present a methodology that uses AGN clustering to set new and valuable \emph{independent} constraints on the overall shape of the local black hole scaling relations. As we describe below, \emph{simultaneously} matching the large-scale AGN clustering and the local black hole mass density provides more robust constraints on the mean black hole radiative efficiency \epsi.

\section*{The connection between AGN clustering, scaling relations and radiative efficiency}
\label{sec|BasicIdea}
Within the framework of a cold dark matter Universe, more massive host dark matter haloes appear progressively more strongly clustered, i.e., their spatial distribution shows more marked departures from an underlying random distribution\cite{CooraySheth}. Galaxies and black holes residing in more massive dark matter haloes are thus naturally expected to appear more clustered. Our methodology builds on this basic notion of galaxy clustering and it can be briefly outlined as follows:
\begin{itemize}
  \item At any redshift of interest $z$, we first create large catalogues of host dark matter haloes from the halo mass function $n(\mhe)$. In practice, in a given (large) volume $V$ we select the halo masses of mass \mh\ for which the cumulative halo mass function is an integer, i.e. $V\times n(>\mhe)=N$, with $N\in \mathbb{I}$. To each dark matter halo we then assign a central galaxy with stellar mass given via the $\mstare[z]-\mhe[z]$ relation. The latter relation is inferred from ``abundance matching'' arguments, i.e., based on number density equivalence between galaxy and host halo number counts\cite{Shankar17g,Grylls19}, $n(>\mstare,z)=n(>\mhe,z)$. We include a scatter of $0.15$ dex in stellar mass at fixed host halo mass\cite{Shankar17g}.
  \item To each ``mock'' galaxy we then assign a central supermassive black hole with mass as given by an input $\mbhe[z]-\mstare[z]$ empirical relation. We include a scatter of $\sim 0.4-0.5$ dex in black hole mass at fixed stellar mass\cite{Shankar16BH,Shankar19BH}, as detailed below. We will also explore a model in which we bypass the $\mbhe[z]-\mstare[z]$ relation assigning black holes via the $\mbhe-\sise$ relation (with a scatter of $0.3$ dex in black hole mass), as explained below.
  \item The combination between black-hole and galaxy property (stellar mass or velocity dispersion), and galaxy property with host halo mass effectively predicts a $\mbhe[z]-\mhe[z]$ relation.
  \item More massive haloes are more strongly clustered, and their (large-scale) clustering strength is encoded in a parameter ``b'' called the bias, as explained below. More massive haloes are on average characterized by larger $b$-values.
  \item For a given input $\mbhe[z]-\mstare[z]$ relation, we compute the implied $\mbhe[z]-\mhe[z]$ relation, and finally the predicted large-scale clustering of black holes, encoded in the $b-\mbhe$ relation.
    \item The higher the normalization in the $\mbhe[z]-\mhe[z]$ relation, the lower is the $\mhe$ that hosts a given $\mbhe$,
  and so the weaker is the expected clustering strength of black holes at fixed black hole mass (and dispersion around the mean).
  \item The $\mbhe[z]-\mhe[z]$ relation(s) that provide the closest match to the AGN clustering measurements, will set key constraints on, most noticeably, the most appropriate input $\mbhe[z]-\mstare[z]$ relation.
  \item After assigning to each galaxy a supermassive black hole mass via the $\mbhe[z]-\mstare[z]$ relation favoured by AGN clustering, we then estimate the implied $V_{\rm max}$-weighted black hole mass density and mass function.
  \item Finally, the match between the local mass density and accreted mass density determined from integrated AGN number counts, will yield more secure constraints on the mean radiative efficiency of supermassive black holes.
\end{itemize}

\section*{Results}

\subsection{The local \mbh-\mstar\ relation and implied \mbh-\mh\ relation}
\label{subsec|compareLocalMbhMstar}

\begin{figure*}
\begin{center}
    \center{\includegraphics[width=\textwidth]{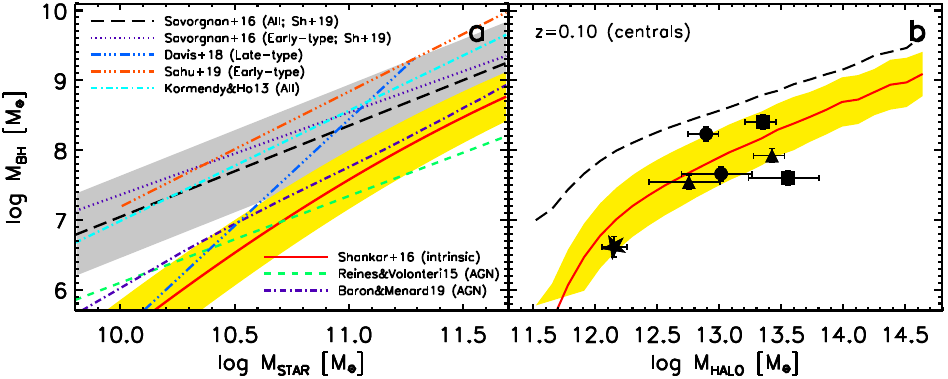}
    \caption{\textbf{Overview of local scaling relations between black hole mass and host (total) stellar mass.}
    \emph{Left}: Correlations between central black hole mass and host galaxy total stellar mass in the local Universe. The black long-dashed line with its scatter (grey band) is the relation inferred from the (whole) quiescent sample by Savorgnan et al.\cite{Savorgnan16}, while the purple, dotted line refers to only their early-type galaxies (ellipticals and lenticulars). The triple dot-dashed orange line is the fits to local quiescent samples of early-type galaxies with dynamical measures of black holes by Sahu et al.\cite{Sahu19}. The dot-dashed, cyan line is a linear fit to the sample of Kormendy \& Ho\cite{KormendyHo}. The solid red line with its scatter (yellow region) is the unbiased \mbh-\mstar\ relation from Shankar et al.\cite{Shankar16BH}. The green dashed and purple dot-dashed lines are the fits to the local active samples from, respectively, Reines \& Volonteri\cite{ReinesVolonteri15} and Baron \& M\'{e}nard\cite{Baron19}, while the blue triple-dot line is the fit to a local sample of late-type galaxies by Davis et al.\cite{Davis18} with dynamically-measured black hole masses. \emph{Right}: Correlations between black hole mass host halo mass at $z=0.1$ (halos are defined to be 200 times the background density) implied by the \mstar-\mhalo\ relation extracted from abundance matching coupled to the \mbh-\mstar\ observed and unbiased relations (black long-dashed and red solid lines in the left panel). Data are from Powell et al.\cite{Powell18} (squares), Krumpe et al.\cite{Krumpe15,Krumpe18}(triangles and circles, respectively) and the Milky Way black hole\cite{Ghez08,Posti19MW} (star).
    \label{fig|FigureMbhMstar}}}
\end{center}
\end{figure*}

We start by discussing the \mbh-\mstar\ relation between black hole mass and host galaxy (total) stellar mass, which is perhaps the most actively debated correlation between black holes and galaxies (\figu\ref{fig|FigureMbhMstar}) yet widely adopted as a reference in black hole-galaxy co-evolution cosmological models\cite{Dave19}.
The left panel of \figu\ref{fig|FigureMbhMstar} reports the latest renditions of the \mbh-\mstar\ relation (full details on the different scaling relations can be found in the ``Methods' Section) from local galaxy samples with dynamically-measured black hole masses from Savorgnan et al.\cite{Savorgnan16} (for their subsample of secure black-hole mass measurements\cite{Shankar16BH,Shankar19BH}),  Kormendy \& Ho\cite{KormendyHo}, and Sahu et al.\cite{Sahu19}, as labelled in the top left corner. The same panel also includes, for completeness, the \mbh-\mstar\ relation (blue, triple dot-dashed line) recently derived by Davis et al.\cite{Davis18} for a local sample of 40 spiral galaxies, which appears to be around three times steeper than the one characterizing the observed sample of early-type galaxies (dotted line). Also shown is the \mbh-\mstar\ relation from Reines \& Volonteri\cite{ReinesVolonteri15} (dashed, green line) and Baron \& M\'{e}nard\cite{Baron19} (dot-dashed purple line) extracted, respectively, from a local AGN samples of 244 and 2000 sources with single-epoch black hole mass estimates. The ``intrinsic'' \mbh-\mstar\ relation proposed by Shankar et al.\cite{Shankar16BH} is reported by a solid red line with its scatter marked by the yellow region. Hereafter, we will refer to the de-biased \mbh-\mstar\ relation put forward by Shankar et al.\cite{Shankar16BH} (solid red line in the left panel of \figu\ref{fig|FigureMbhMstar}, with a scatter of $\lesssim 0.4$ dex, marked by the yellow band) as the ``intrinsic'' or ``unbiased'' scaling relation.

The very first feature to highlight in the left panel of \figu\ref{fig|FigureMbhMstar} is the systematic discrepancy between the fits to the local quiescent samples of supermassive black holes and Shankar et al.'s intrinsic \mbh-\mstar\ relation. Shankar et al.\cite{Shankar16BH} devised targeted Monte Carlo simulations to show that local quiescent (mainly early-type) galaxies having dynamically measured black hole mass, present larger velocity dispersions at fixed stellar mass with respect to galaxies in the Sloan Digital Sky Survey (SDSS). In fact, given the strong and fundamental\cite{Bernardi07,Shankar16BH,Shankar17BH,Shankar19BH} dependence $\mbhe\propto \sise^{4-5}$, higher velocity dispersions
would on average select higher mass black holes at fixed host galaxy stellar mass. Thus, the local quiescent samples of (mainly early-type) galaxies with dynamically measured black holes would then follow \emph{biased} scaling relations, with fictitiously higher normalizations than the bulk of the underlying population of supermassive black holes. To a large extent, Shankar et al.\cite{Shankar16BH} showed that a possible reason behind the higher velocity dispersions in the local sample of quiescent early-type galaxies could be ascribed to a bias arising from the requirement that the central black hole's sphere of influence must be resolved to measure black hole masses with spatially resolved kinematics.

As already noted by Reines \& Volonteri, the relations defined by the AGN samples are about an order of magnitude lower than the ones defined by the quiescent early-type samples, in line with several other previous claims\cite{Sarria10,busch2014low,Falomo14}. In fact, Shankar et al. (2019)\cite{Shankar19BH} have highlighted the fact that the vast majority of local AGN samples follow relations that are noticeably lower in normalization than those of quiescent galaxies and closer to the intrinsic relation (yellow region). They argue that AGN do not suffer from angular resolution-selection effects, so they may more faithfully trace the intrinsic black hole-galaxy scaling relations.

In summary, different host galaxy morphology, observational biases and selection effects all tend to result in different black hole mass--host galaxy scaling relations.  To make progress, we need an independent constraint on these scaling relations, which we obtain as follows.

It is well known that galaxies with large stellar masses are more strongly clustered\cite{CooraySheth}. Therefore, it should be possible to set constraints on the normalization and steepness of the \mbh-\mstar\ relation by including \emph{independent} galaxy clustering measurements. The right panel of \figu\ref{fig|FigureMbhMstar} reports the $z=0.1$ implied \mbh-\mh\ relation of central galaxies obtained by assigning to each dark matter halo extracted from the parent halo mass function\cite{Tinker08}, a galaxy stellar mass from the mean \mstar-\mh\ relation (and its scatter), and then a black hole mass from the \mbh-\mstar\ relation (with scatter) for the quiescent Savorgnan et al.\cite{Savorgnan16} sample (black long-dashed line) and the intrinsic \mbh-\mstar\ relation from Shankar et al. (2016; solid red line and its scatter marked by the yellow region)\cite{Shankar16BH}. The latter \mbh-\mstar\ relation clearly predicts a correlation between black hole mass and host halo mass that is more consistent with the variety of data extracted from low-redshift AGN large-scale optical surveys\cite{Krumpe15,Krumpe18,Powell18} and reported in the right panel of \figu\ref{fig|FigureMbhMstar}. In what follows we always adopt as a reference for the ``biased/observed'' \mbh-\mstar\ relation the one inferred from the Savorgnan et al. sample. As inferred from the left panel of \figu\ref{fig|FigureMbhMstar}, the Savorgnan et al. relation is the shallowest and the lowest in normalization among the various renditions of the dynamically-based \mbh-\mstar\ scaling relations.
Adopting any of the other \mbh-\mstar\ relation characterizing quiescent black hole samples\cite{Sahu19} would, if anything, further exacerbate the discrepancy with AGN clustering measurements.

\subsection{Independent constraints on the \mbh-\mstar\ relation from AGN large-scale clustering}
\label{subsec|compareClustering}

\begin{figure*}
\begin{center}
    \center{\includegraphics[width=\textwidth]{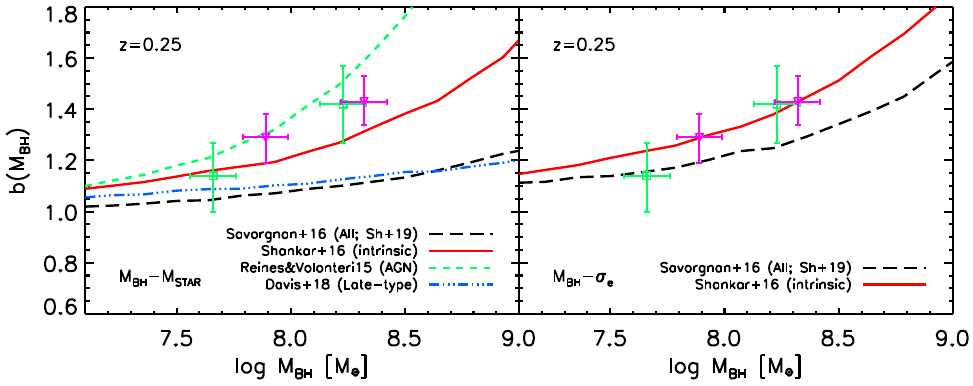}
    \caption{\textbf{Predicted bias as a function of black hole mass.} Left: Results for the mean large-scale clustering are shown at $z=0.25$ for unbiased and observed (red solid and long-dashed black lines, respectively), the Reines \& Volonteri\cite{ReinesVolonteri15} (dashed green line), and the (late-type) Davis et al.\cite{Davis18} (blue triple-dot-dashed line) \mbh-\mstar\ relations, as labelled. The data\cite{Krumpe15} (green squares and purple triangles are X-ray and optical AGN, respectively) are extracted from the clustering properties of AGN identified in ROSAT and SDSS in the redshift range $0.16<z<0.36$. The models with the unbiased/lower normalization \mbh-\mstar\ relation are favoured by current data. Right: Similar to left hand panel, but now showing $b$(\mbh) expected from the observed/biased (dashed) and intrinsic (solid) \mbh-\sis\ scaling relations.
    \label{fig|FigureBias}}}
\end{center}
\end{figure*}

To further improve on the constraints that can be derived from AGN samples, we now use the large-scale AGN clustering measurements performed on the ROSAT All-Sky Survey and SDSS data sets\cite{Krumpe15} in the redshift range $0.16<z<0.36$. AGN clustering is usually quantified by the two-point correlation function:  the excess number of AGN pairs in the data compared to a random distribution of the same number density, as a function of separation. At scales larger than a few Mpc, we are in the linear regime of structure formation and the ratio of the two-point correlation function to that expected for the dark matter is approximately independent of pair separation.  The square root of this ratio is called the ``large-scale bias factor'' $b$, and it is not to be confused with the bias in the scaling relations we were discussing about in the previous Sections!  The large-scale bias factor $b$ differs for different galaxy samples. The bias factor in fact progressively increases for larger halo masses\cite{CooraySheth}. Due to the monotonic dependence between stellar mass and host halo mass in the \mstar-\mh\ relation, especially at the masses of interest in this work (intrinsic scatter of $0.15$ dex), samples with larger \mstar, will on average appear more strongly clustered and characterized by a larger bias factor $b$\cite{CooraySheth}. Since black hole mass is also monotonically related to host halo mass (albeit with a larger scatter than in the \mstar-\mh\ relation), we would thus expect the $b$ factor to increase with black hole mass.

The two green squares with error bars in \figu\ref{fig|FigureBias} (same in both panels) show the large-scale bias factors for two samples of low- and high-mass active supermassive black holes observed in X-ray band. Similarly, the two magenta triangles show the bias factor $b$ in a low- and high-mass sample of optically-selected AGN.
The inferred bias factors $b$ appear to be rather independent of wavelength, and indeed larger for the samples with larger black hole mass \mbh.

The curves in the left-hand panel of \figu\ref{fig|FigureBias} show the expected large-scale bias assuming black hole masses are assigned via the observed/biased (long-dashed black, \eq\ref{eq|ObservedMbhMstar}) or intrinsic/unbiased (solid red) \mbh-\mstar\ scaling relations\cite{Shankar16BH} (\eqs\ref{eq|IntrinsicMbhMstar} and \ref{eq|IntrinsicScatterMbhMstar}), with the same underlying mean \mstar-\mhalo\ relation (see the ``Methods'' Section). The predicted bias parameter $b$ based on the observed/biased scaling relations is too low and is ruled out at high confidence \textbf{($\gtrsim 3\sigma$)}, whilst that based on the intrinsic/unbiased scaling relations provides a good match to the data. The agreement is remarkable, given that there are no free parameters to fit.  In the Methods Section we provide a detailed comparison with the measured cross-correlation of AGN-SDSS sources and careful estimates of the significance of the discrepancies between models and data. The dashed green curve shows that the Reines \& Volonteri\cite{ReinesVolonteri15} \mbh-\mstar\ scaling relation also yields a good match to the clustering measurements, implying that the AGN clustering measurements are still consistent with \mbh-\mstar\ relations that are even lower in normalization than Shankar et al.'s intrinsic relation, further strengthening the key results of this work. For the Reines \& Volonteri relation we assume a nominal intrinsic scatter smaller of 0.3 dex, of the order of what inferred by the Authors, but slightly smaller than the ones characterizing the scaling relations of dynamically-measured inactive black holes\cite{Shankar19BH}.

It has been argued that host galaxy morphology may play a substantial role in creating the systematic offset between black hole scaling relations of active and quiescent galaxies. In fact the former active black holes could be mostly hosted in later-type galaxies, which are expected to allegedly be hosting lower black hole masses at fixed stellar mass, while the latter quiescent samples with larger black hole masses tend to inhabit earlier-type galaxies\cite{ReinesVolonteri15}. However, the Davis et al.\cite{Davis18} steep relation of dynamically-measured quiescent black holes in late-type galaxies (blue triple dot-dashed line, with an intrinsic scatter of 0.7 dex), tends to still fall substantially below the bias data at high \mbh. The latter result clearly supports the view of a bias in the observed \mbh-\mstar\ relation which is independent of galaxy morphology or AGN type. We also expect the Krumpe et al. AGN clustering measurements, based on large, serendipitous samples, to be representative of a variety of host galaxy morphologies and not just late types. We finally note that all of the data in \figu\ref{fig|FigureBias} mostly refer to Type 1 AGN. Observationally, Type 2/obscured AGN have anyhow been measured to cluster at a comparable or even higher level, at least at $z<1$\cite{DiPompeo16,Jiang16bias}, which would further strengthen our results in favour of lower normalizations in the \mbh-\mstar\ intrinsic relation.

In summary, the low-redshift AGN clustering measurements considered here favour considerably \emph{lower} normalizations of the \mbh-\mstar\ relation.
Scalings in \mbh-\mstar\ with high normalizations such as the observed/biased one characterizing quiescent galaxies (black long-dashed line), tend to map black holes to lower stellar masses and thus to lower host halo masses (see \figu\ref{fig|FigureMbhMstar}), significantly decreasing the predicted clustering signal. On the other hand, \mbh-\mstar\ scalings such as those from the intrinsic relation (red solid line) or from the AGN sample (green dashed line), better line up with the clustering data. AGN clustering thus provides additional, independent evidence for the presence of a bias in the observed \mbh-\mstar\ relation, at least for quiescent black holes with $\mbhe \gtrsim 3\times 10^7\, \msune$. At lower black hole masses AGN clustering data that are currently available lose their constraining power. For example, for $\mbhe \lesssim 3\times 10^7\, \msune$, AGN clustering data are not able to distinguish between the steep (late-type) relation of Davis et al. or the flatter one of Reines \& Volonteri.  This loss in constraining power is a consequence of the fact that $b$(\mhalo) becomes nearly constant at low masses\cite{ShethTormen}.

For completeness, the right-hand panel of \figu\ref{fig|FigureBias} shows the bias parameter $b$(\mbh) expected from the \mbh-\sis\ scaling relation. Measuring the clustering strength implied by the \mbh-\sis\ relation represents an additional independent test to the overall reliability of the black hole scaling relations. This test is also particularly valuable as it circumvents the use of \mstar\ by directly applying abundance matching between the velocity dispersion-based local black hole mass function and halo (plus subhalo) mass functions (see the ``Methods" \sect\ for full details). Moreover, it has been claimed that the \mbh-\sis\ relation is less biased than the \mbh-\mstar\ relation, implying that the former should be closer to the ``intrinsic'' relation, in terms of normalization, slope, dispersion around the mean\cite{Bernardi07,Shankar16BH,Shankar19BH}. It is thus expected that the related (large-scale) clustering properties should be less sensitive to observational biases. Indeed, when adopting reasonable scatter ($\sim 0.3$ dex) around the mean \mbh-\mhalo\ relation\cite{Shankar10shen,Aversa15}, when adopting the observed (dashed black) and intrinsic (solid red) \mbh-\sis\ relations, in both cases we find mean $b$-\mbh\ relations broadly consistent with the data, with the former only slightly disfavoured at the highest black hole masses.

\subsection{Comparing local and accreted mass functions: Constraining the mean radiative efficiency of supermassive black holes}
\label{subsec|ContinuityEquation}

\begin{figure*}
\begin{center}
    \center{\includegraphics[width=\textwidth]{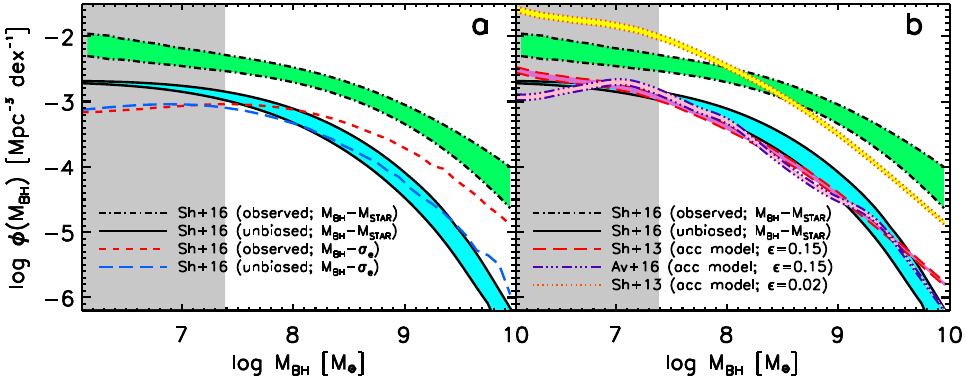}
    \caption{\textbf{Comparing local and accreted mass functions.} \emph{Left}: Estimates of the local black hole mass functions as derived from the observed \mbh-\mstar\ relation (dot-dashed lines and green region), the unbiased\cite{Shankar16BH} \mbh-\mstar\ relation (solid black lines and cyan region), the observed \mbh-\sis\ relation (red dashed line), and the unbiased\cite{Shankar16BH} \mbh-\sis\ relation (blue long-dashed line). A mismatch between the observed \mbh-\mstar- and \mbh-\sis-based black hole mass functions (green region and red dashed line, respectively) is evident. \emph{Right}: Comparison between the local black hole mass functions from the left panel with the accreted black hole mass functions at $z=0.1$ from the Shankar et al.\cite{Shankar13acc} and Aversa et al.\cite{Aversa15} accretion models (dotted red and purple triple dot-dashed lines, respectively; the coloured areas in between the lines, yellow and pink respectively, represent the uncertainty region from excluding/including Compton-thick AGN with $25<\log N_H/{\rm cm^{-2}}<26$). Irrespective of the input Eddington ratio distributions, both accretion models, which use a radiative efficiency of $\epsie=0.15$, are in good agreement with the unbiased local black hole mass function (cyan region). The grey areas in both panels mark the regions in which local measurements are less robust and cannot be constrained by clustering data.
    \label{fig|FigureBHMF}}}
\end{center}
\end{figure*}

\begin{figure*}
\begin{center}
    \center{\includegraphics[width=\textwidth]{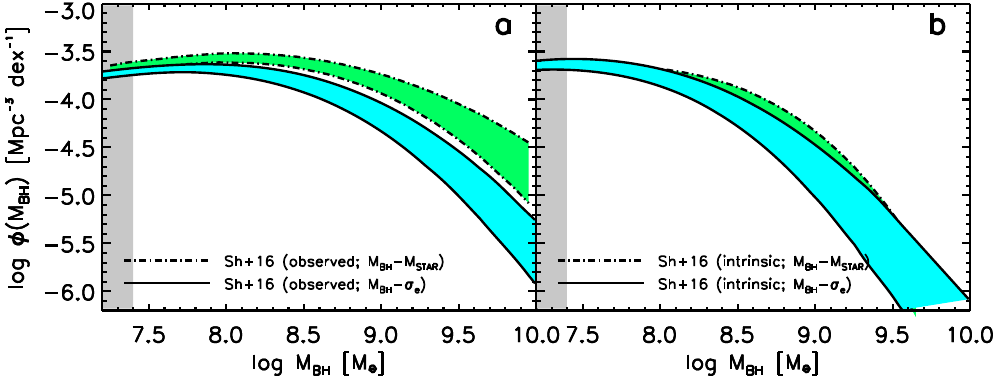}
    \caption{\textbf{Predicted local black hole mass functions of elliptical galaxies.}
    Same format as \figu\ref{fig|FigureBHMF} but only for Elliptical, bulge-dominated galaxies. Lef/right panel show the comparison between the observed/intrinsic \mbh-\mstar\ and \mbh-\sis\ relations of early-type galaxies, as labelled.
    \label{fig|BHMFsEll}}}
\end{center}
\end{figure*}

We first compute the local mass function of supermassive black holes from the SDSS galaxy sample adopted by Shankar et al.\cite{Shankar16BH} (full details given in the Methods Section). The 1$\sigma$ error in number densities at fixed black hole mass are shown in \figu\ref{fig|FigureBHMF} for the unbiased and biased scaling relations, respectively. The left panel of \figu\ref{fig|FigureBHMF} points to a large difference between black hole mass functions computed from the observed and intrinsic \mbh-\mstar\ scaling relations (green and cyan areas, respectively; the vertical grey band marks the region in which black hole mass function estimates become progressively less secure). Of uttermost importance, the black hole mass function derived from the observed \mbh-\sis\ relation (dashed red line) still presents noticeable departures from the one derived from the observed \mbh-\mstar\ relation (green region). As proven in \figu\ref{fig|BHMFsEll}, this behaviour is \emph{not} a consequence of neglecting bulge corrections but it reflects an internal inconsistency in the samples of local quiescent black holes\cite{Bernardi07,Shankar16BH,Shankar19BH}. On the other hand, the black hole mass function derived from the intrinsic \mbh-\sis\ relation\cite{Shankar16BH} (long-dashed blue line in \figu\ref{fig|FigureBHMF}) is fully consistent with the black hole mass function derived from the intrinsic \mbh-\mstar\ relation (cyan region; \eq\ref{eq|IntrinsicMbhMstar}).

In the right panel of \figu\ref{fig|FigureBHMF} we then compare the local black hole mass functions with the
accreted mass functions derived from AGN number counts and continuity equation techniques. 
We show results for the accreted black hole mass function computed in two ways, via the method outlined in Shankar et al.\cite{Shankar13acc} (long-dashed red lines) and as proposed by Aversa et al.\cite{Aversa15} (triple dot-dashed purple lines). For both models we adopt the same input X-ray AGN luminosity function\cite{Ueda14} with bolometric correction from Yang et al.\cite{Yang18}, and a mean radiative efficiency of $\epsie=0.15$. In all accretion models we include obscured AGN with column densities up to $\log N_H/{\rm cm^{-2}}\lesssim 25$, as observed in a number of studies\cite{Harrison16}. The two sets of lines for each model bracket the systematic uncertainty in including or excluding obscured AGN with higher column densities, $25<\log N_H/{\rm cm^{-2)}}<26$, following the $N_H$ column density distributions provided by Ueda et al.\cite{Ueda14}. Despite the substantially different input Eddington ratio distributions, both accretion models predict similar accreted black hole mass functions that line up well with the local black hole mass function derived from the intrinsic \mbh-\mstar\ relation (cyan area). Broadly matching the local black hole mass function derived from the observed \mbh-\mstar\ relation would instead require at least one order of magnitude lower radiative efficiencies (dotted red lines and yellow area). It is interesting to note the good match between local and accreted mass function even at the largest mass scales. As discussed in previous works\cite{Shankar13acc,Aversa15}, frequent black hole mergers tend to generate a relatively long high-mass tail in the accreted black hole mass function. The latter would be in tension with the nearly-exponential fall off of the local black hole mass function, unless an appropriately fine-tuned radiative efficiency progressively increasing with black hole mass is adopted\cite{Shankar13acc}.

\begin{figure*}
\begin{center}
    \center{\includegraphics[width=\textwidth]{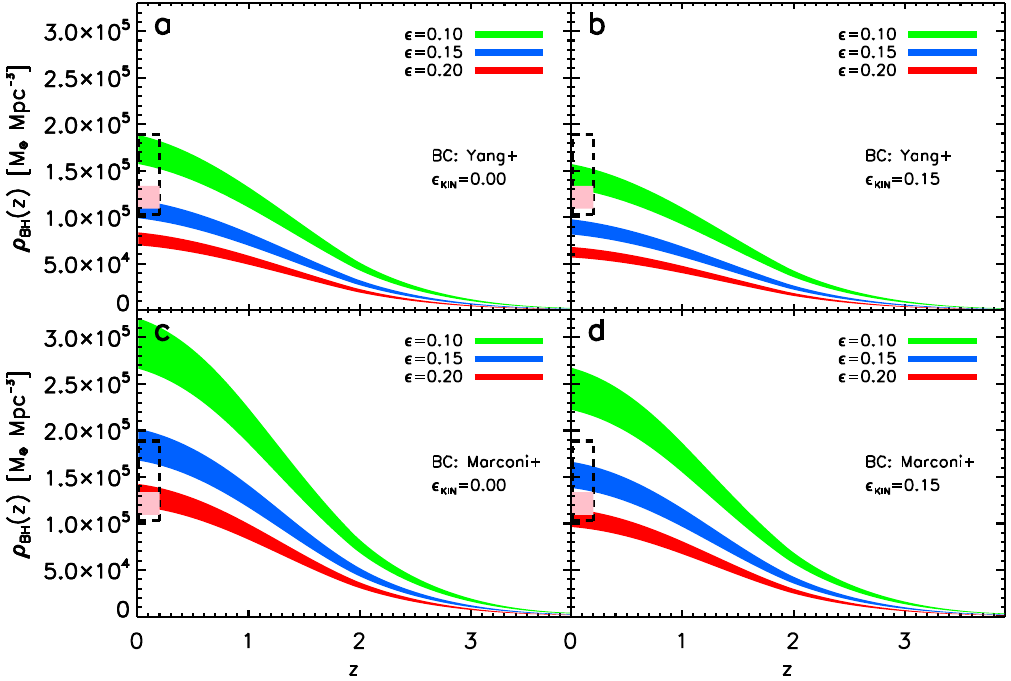}
    \caption{\textbf{Comparing local and accreted integrated mass densities.} ``Relic'' black hole mass density $\rho_{\rm SMBH}$ inferred from the local unbiased\cite{Shankar16BH} \mbh-\mstar\ and \mbh-\sis\ relations (black dashed box and solid pink square, respectively), compared with the accreted mass density $\rho_{\rm AGN}$ in AGN assuming different values of the radiative efficiency $\epsie$, bolometric correction BC, and kinetic efficiency $\epsie_{\rm kin}$, as labelled. The coloured region for each model shows the uncertainty from excluding/including Compton-thick AGN with $25<\log N_H/{\rm cm^{-2)}}<26$. For the range of currently accepted values of $\epsie_{\rm kin}$, BC and  obscured fraction, the data indicate moderate-to-high radiative efficiencies ($\epsie\gtrsim 0.1$).
    \label{fig|FigureRhoz}}}
\end{center}
\end{figure*}

In integral form, \figu\ref{fig|FigureRhoz} shows the comparison between the local black hole and accreted mass \emph{densities}. The former is obtained from direct integration of the local black hole mass function derived from the \mbh-\mstar\ relation (long-dashed black box, corresponding to integration of the cyan region in \figu\ref{fig|FigureBHMF}; also included, for completeness, the mean local black hole mass density implied by the intrinsic \mbh-\sis\ relation, pink square). The latter accreted mass density is retrieved from integration of the AGN bolometric luminosity function over luminosity and cosmic time (see Methods Section). The three coloured bands in each panel of \figu\ref{fig|FigureRhoz} mark three different values of the radiative efficiency, as labelled in each panel. Each coloured band brackets the systematic uncertainty induced by including or neglecting the Compton-thick AGN with $25<\log N_H/{\rm cm^{-2}}<26$. The top row includes model outputs that make use of the Yang et al.\cite{Yang18} and Marconi et al.\cite{Marconi04} bolometric corrections, respectively. The right panels also allow for an additional loss of rest-mass energy via, e.g., jets and/or winds, parameterized by the kinetic efficiency which we set\cite{Shankar08Cav} to an average value of $\epsie_{\rm kin}=0.15$. It is interesting to note that Yang et al.\cite{Yang18} also computed the redshift-dependent \mbh-\mstar\ relation from integrating the AGN X-ray emissivity at different epochs along (putative) stellar mass accretion histories. By assuming a constant mean radiative efficiency of \epsi=0.1 to convert from AGN bolometric luminosities to black hole accretion rates, they claimed a local \mbh-\mstar\ relation in close agreement to the unbiased one by Shankar et al.\cite{Shankar16BH}, and definitely steeper than the one from Kormendy \& Ho\cite{KormendyHo}.

It is clear from \figu\ref{fig|FigureRhoz} that, irrespective of the exact assumed value of the obscured fraction, bolometric correction or kinetic efficiency, all accretion models always require relatively high radiative efficiencies $\epsie\sim 0.10-0.20$ to match the local black hole mass density. This yields a mean radiative efficiency a factor of $\sim 2-3$ higher than previous estimates solely based on observed/biased black hole scaling relations\cite{Marconi04,Ueda14}. Our results would support black hole spin parameters\cite{Bardeen72} $a \gtrsim 0.7$, which are consistent with a number of direct measurements performed via X-ray reflection analysis\cite{Reynolds19} or spectral energy distribution fits\cite{Shankar16,ZhangLu19}. Hints for relatively high mean radiative efficiencies are also present in some previous works\cite{Elvis02,YuLu08}. However those studies adopted biased estimates of the local black hole mass density, sometimes \textbf{relatively} crude estimates of AGN emissivities, and without the vital constraints from AGN clustering. In this work we have ultimately shown that the AGN large-scale clustering as a function of black hole mass represents a fundamental discriminator for black hole-galaxy scaling relations. Upcoming large-scale AGN surveys such as eROSITA\cite{Merloni_eRosita}, probing millions of AGN, will generate exquisite measurements of AGN clustering, set stringent constraints of the scaling relations down to the lowest accessible black hole masses, and in turn tighter constraints on the radiative and kinetic efficiencies of accreting supermassive black holes.

\noindent \textbf{References}\\
\vspace{0.3cm}
\bibliographystyle{naturemag}

\begin{methods}


\subsection{Details on the different \mbh-\mstar\ relations.}
\label{subsec|MethodsFits}

The left panel of \figu\ref{fig|FigureMbhMstar} shows a number of linear fits to the \mbh-\mstar\ relation from different groups and black hole samples. Following Shankar et al.\cite{Shankar19BH}, we first fitted the subsample of dynamical masses from Savorgnan et al.\cite{Savorgnan16} with reliable black hole mass measurements. We adopted the Bayesian-based linear fitting IDL routine \texttt{$LINMIXERR$} to compute all median slopes and normalizations. We verified that the variance-based fitting prescriptions reported in Appendix A of Shankar et al.\cite{Shankar17BH} provide identical mean values. The (intrinsic) scatter was instead computed as $\sigma=\sqrt{y_{\rm rms}^2-\left\langle \Delta(y_i) \right\rangle^2}$, where $y_{\rm rms}^2=\sum_i^N \left[y_i-y(x)\right]^2 / (N-1)$ with $y(x)$ the linear best-fit, and $\Delta(y_i)$ are the quoted uncertainties in black hole masses. The resulting fit to the (full) Savorgnan et al. sample (long-dashed black line) reads as
\begin{equation}
\log \frac{\mbhe}{\msune} =
 8.35\pm0.09 + (1.31\pm0.22)\times \log \left(\frac{\mstare}{10^{11}\msune}\right)\, .
 \label{eq|ObservedMbhMstar}
\end{equation}
with an intrinsic scatter of $0.5\pm0.1$ dex.

The zero points of the correlations from Davis et al.\cite{Davis18} and Sahu et al.\cite{Sahu19}, their \eqs3 and 11, respectively, have been modified setting the parameter $v$ in their fits equal to $v=0.64$, to allow for conversion to the mass-to-light ratios of Bell et al.\cite{Bell03SEDs} with a Chabrier\cite{Chabrier03} stellar initial mass function adopted as a reference in Shankar et al. (2016)\cite{Shankar16BH} and in this work. The Reines \& Volonteri fit\cite{ReinesVolonteri15} and the Kormendy \& Ho\cite{KormendyHo} sample, which is also taken from Table 3 of Reines \& Volonteri, have been corrected for the different mass-to-light ratios following \eq A1 in Shankar et al. (2019)\cite{Shankar19BH}. A direct linear fit to the (corrected) Kormendy \& Ho\cite{KormendyHo} sample retrieved from Reines \& Volonteri yields $\log \frac{\mbhe}{\msune} = 8.56 + 1.58\times \log \left(\frac{\mstare}{10^{11}\msune}\right)$ (dot-dashed cyan line in the left panel of \figu\ref{fig|FigureMbhMstar}). The sample by Baron \& M\'{e}nard\cite{Baron19} has been converted to the Bell et al. mass-to-light ratios adopting the mass-dependent correction presented in \figu A2 of Bernardi et al. (2017)\cite{Bernardi17}.

Last but not least, in the same left panel of \figu\ref{fig|FigureMbhMstar} we report the intrinsic (or unbiased/de-biased) correlation put forward by Shankar et al. (2016)\cite{Shankar16BH}, which is the result of a series of Monte Carlo simulations.
The mean intrinsic relation by Shankar et al. between black hole mass and total host galaxy stellar mass \mbh-\mstar\ is well approximated by the relation
\begin{multline}
\log \frac{\mbhe}{\msune} =
 7.574 + 1.946\,\log \left(\frac{\mstare}{10^{11}\msune}\right)- 0.306\\
\times\left[\log\left(\frac{\mstare}{10^{11}\msune}\right)\right]^2
 - 0.011\,\left[\log\left(\frac{\mstare}{10^{11}\msune}\right)\right]^3\, ,
 \label{eq|IntrinsicMbhMstar}
\end{multline}
with a mass-dependent intrinsic scatter in (the logarithm of) black hole mass at fixed stellar mass (yellow region) well reproduced by\cite{Sesana16}
\begin{equation}
\Delta \log \frac{\mbhe}{\msune} =
 0.32 - 0.1\times \log \left(\frac{\mstare}{10^{12}\msune}\right)\, .
 \label{eq|IntrinsicScatterMbhMstar}
\end{equation}
Both \eqs\ref{eq|IntrinsicMbhMstar} and \ref{eq|IntrinsicScatterMbhMstar} are good fits\cite{Shankar16BH,Shankar19BH} to galaxies with stellar mass above a few times $\log \mstare/\msune>10$.
As recently discussed by Shankar et al. (2019)\cite{Shankar19BH}, \eq\ref{eq|IntrinsicMbhMstar} was obtained by imposing that black holes closely follow, via the $\mbhe \propto \sise^{4-5}$ relation, the velocity dispersion-stellar mass of SDSS early-type galaxies.
Any significant contribution from later-type galaxies would tend, if anything, to produce slightly lower normalizations of the unbiased \mbh-\mstar, thus further increasing the mismatch with the scaling relation of quiescent galaxies.

\subsection{Abundance matching technique.}
\label{subsec|AbundanceMatching}

The \mstar-\mh\ relation is obtained from abundance matching between the stellar mass and the halo plus subhalo mass functions\cite{Shankar17g,Shankar18}. We here adopt the Tinker et al. (2008) halo mass function\cite{Tinker08} and the unevolved, surviving subhalo mass function by Jiang and van den Bosch\cite{Jiang16SubMF} (see their Section 4.5). The latter is the relevant statistical description of subhaloes that survived mergers down to a given redshift, and thus the most suited quantity to include in accurate abundance matching algorithms. However, we verified that adopting different, more general fits of the subhalo mass function\cite{Giocoli08} would yield very similar results.

We assume throughout an intrinsic scatter of 0.15 dex in stellar mass at fixed halo mass\cite{Shankar17g}. We include scatter following \eq5 in Aversa et al.\cite{Aversa15}, though very similar results are found by following the methodology presented in Shankar et al. (2017)\cite{Shankar17g}. We adopt the S\'{e}rsic+Exponential stellar mass function from Bernardi et al. 2013\cite{Bernardi13}, which is based on the same photometry and mass-to-light ratios used by Shankar et al. (2016) and in this work. We assume no evolution in the stellar mass function up to $z=0.25$, which is a very good approximation according to the latest estimates\cite{Bernardi16,Tinker17}.

The \mstar-\mhalo\ relation is now more securely established at least for relatively high stellar masses $\mstare \gtrsim 3\times 10^{10}\, \msune$ and at relatively low redshifts\cite{Bernardi13,Kravtsov18,Shankar17g,Behroozi18,Moster18,Grylls19}. We then extract large host halo catalogues from the halo plus subhalo mass function and assign galaxies to haloes via the \mstar-\mhalo\ relation. We note that all central and satellite galaxies are here assigned from the \mstar-\mhalo\ relation at the single redshift of $z=0.25$, which is an extremely good approximation to a full multi-epoch abundance matching routine, especially for relatively massive galaxies ($\mstare\gtrsim 3\times 10^{10}, \msune$), as recently demonstrated by Grylls et al.\cite{Grylls19}. In this work we always assume, unless otherwise noted, a standard reference cosmology of $\Omega_m=0.3$, $h=0.7$, $\sigma_8=0.8$, $n_s=1$, as adopted in the observational papers we compare with.

\subsection{Computing local and accreted mass functions.}
\label{subsec|ContinuityEquation}

The local black hole mass function is computed from the SDSS galaxy sample adopted by Shankar et al.\cite{Shankar16BH}. In a Monte Carlo fashion, to each galaxy a supermassive black hole is assigned following a given input \mbh-\mstar\ relation. To allow for moderate systematics in the galaxy stellar mass estimates due to fine details in the light profiles and/or mass-to-light ratios, at each iteration we randomly add $\pm 0.1$ dex to the stellar masses of each host galaxy, and then assign to it a supermassive black hole from the \mbh-\mstar\ relation inclusive of its scatter. At each iteration, the black hole mass function is then computed from the $V_{\rm max}$ weights assigned to each SDSS galaxy. The final 1$\sigma$ error in number density at fixed black hole mass is then given by the dispersion in number densities resulting from 100 realizations. Supplementary Table~1 reports the results of the black hole mass function and its 1$\sigma$ uncertainty at fixed black hole mass derived from the \mbh-\mstar\ relation. The black hole mass function implied by the \mbh-\sis\ relation is fully consistent with the one given in Supplementary Table~1 within the uncertainties, as shown in \figu\ref{fig|FigureBHMF}.

In Supplementary \figu\ref{fig|BHMFsEll} we show the results of applying the Monte Carlo methodology described above to only elliptical galaxies, which we select from the SDSS sample imposing a Bayesian classification probability\cite{Huertas11} $P[E]>0.5$ (our results do not depend on the exact value of this threshold). The left panel shows the comparison between the black hole mass functions implied by the observed \mbh-\mstar\ and \mbh-\sis\ relations of early-type galaxies\cite{Shankar19BH}. A clear mismatch of the order of $\sim 2-3$ in integrated black hole mass density, is evident between the black hole mass functions, proving that the difference in black hole mass estimators is not a simple byproduct of including or not bulge corrections in the calculation. On the other hand, the right panel of \figu\ref{fig|BHMFsEll} shows the good match between the black hole mass functions implied by the unbiased/intrinsic \mbh-\mstar\ and \mbh-\sis\ relations of early-type galaxies.

To compute the accreted (or ``relic'') black hole mass function we instead solve the continuity equation\cite{SmallBlandford,YuTremaine,Shankar04}
\begin{equation}
\frac{\partial n_{\rm bh}}{\partial
t}(\mbhe,t)=-\frac{\partial (\langle \dot{M}_{\rm bh}\rangle
n_{\rm bh}(\mbhe,t))}{\partial \mbhe}\,,
    \label{eq|conteq}
\end{equation}
where $\langle \dot{M}_{\rm bh}\rangle$ is the mean accretion
rate (averaged over the active and inactive populations) of the
black holes of mass \mbh\ and number density $n_{\rm bh}(\mbhe,t)$ at time $t$.

In the main text we show results for the accreted black hole mass function computed in two ways, via the method outlined in Shankar et al.\cite{Shankar13acc} and as proposed by Aversa et al.\cite{Aversa15}. In the former, \eq\ref{eq|conteq} is solved by assuming\cite{Cao10} an input Eddington ratio
distribution (lognormal) and constantly peaked around an Eddington ratio $\log \lambda=-0.6$, with $\lambda\propto L/\mbhe$) of active black holes convolved with the active mass function to yield the observed AGN luminosity function at any cosmic epoch.
In the latter, following the seminal work by Yu \& Lu\cite{YuLu04}, a light curve is instead assumed in input and the active and total black hole mass functions are then self-consistently derived by direct analytic time integration. Aversa et al. also adopted a slim and thin-disk approximation, in which the radiative efficiency is assumed to depend on the current Eddington ratio, in which the radiative efficiency can be close to ~0.3 during super-Eddington accretion, but rapidly approaching the thin-disc constant value at Eddington/sub-Eddington regimes. Given that the latter is the most typical accretion mode for AGN, the Aversa et al. model reduces to a standard, constant radiative efficiency throughout most of the evolution time.

The normalization and shape of the implied duty cycle, i.e., the ratio between the active and total black hole mass functions, depend on the input Eddington ratio distributions $P(\lambda)$, with $\lambda\propto L/\mbhe$. As extensively discussed by Shankar et al.\cite{Shankar13acc}, assuming, for example, a constant Eddington ratio distribution, gradually generates a very low duty cycle at fixed black hole mass at late epochs, which is largely inconsistent with present data\cite{Goulding10}. On the other hand, an Eddington ratio distribution that drops towards lower values of $\lambda(z)$ at lower redshifts, can provide orders of magnitude higher duty cycles, as the same black hole mass bins are mapped to lower luminosity and more abundant AGN.
We checked that increasing the radiative efficiency from $\epsie\sim 0.05$ to $\epsie\sim 0.2$ does not change these behaviours in the output duty cycle.

\subsection{Accreted mass densities.}
\label{subsec|AccretedMassDensities}

The classical\cite{Soltan,Salucci99,Shankar13review} argument based on integrated mass densities, and inclusive of both radiative and kinetic efficiencies\cite{Shankar08Cav,Ghisellini13,Zubo18}, can be cast as follows
\begin{equation}
        \rho_{\rm AGN}=\int dz \int dL \Phi(L,z)L \frac{(1-\epsie-\epsie_{\rm kin})}{\epsie c^2}=\rho_{\rm SMBH} \, ,
        \label{eq|Soltan}
\end{equation}
in which the integrated (or ``relic'') mass density $\rho_{\rm AGN}$ from accretion over all AGN with number densities $\Phi(L,z)$, bolometric luminosity $L$ and redshift $z$, is compared to the local black hole mass density $\rho_{\rm SMBH}$. The integral in \eq\ref{eq|Soltan} is computed over the luminosity range $42<\log L/\ergse<48$ and from $z=6$. The exact limits of integration do not significantly change the results. The values of bolometric luminosity $L$ in input into \eq\ref{eq|Soltan}, implicitly assume a bolometric correction $\kbe(L)$ that is luminosity dependent, folded in the computation of the bolometric luminosity function $\Phi(L,z)$ derived from Ueda et al.'s X-ray luminosity function.

\subsection{Computing large-scale bias}
\label{subsec|Computebias}

To compare models to clustering data, we first perform abundance matching at the average redshift of $z=0.25$. We then extract large host halo catalogues from the halo (plus subhalo) mass function and assign galaxies to haloes via the \mstar-\mhalo\ relation as described above.

On the general assumption that black hole scaling relations do not depend on environment/host halo mass, to each mock galaxy in the catalogue we assign a black hole mass adopting an input \mbh-\mstar\ relation inclusive of scatter. In principle, the mean large-scale bias for black holes with mass in the range \mbh\ and $\mbhe+d\mbhe$ could be straightforwardly estimated as
\begin{equation}
b(\mbhe)=\frac{1}{N_{\rm bin}}\sum_{i=1}^{N_{\rm bin}} b_h\left[M_{\rm halo,i}(\mbhe)\right]\, ,
    \label{eq|bias}
\end{equation}
where the sum runs over the total number $N_{\rm bin}$ of (central and satellite) parent haloes hosting central black holes with mass in the range \mbh\ and $\mbhe+d\mbhe$.

However, \eq\ref{eq|bias} neglects the probabilities for central and satellite black holes to be active. In the case of AGN in fact, not all galaxies in a given host halo mass bin necessarily contribute to the same clustering signal, and \eq\ref{eq|bias} should be modified to include the duty cycles, or probabilities, $U_{\rm cen}(\mbhe)$ and $U_{\rm sat}(\mbhe)$ of, respectively, central and satellite black holes of mass \mbh\ to be active at a given cosmic epoch. We here follow the same convention as in Shankar et al. (2013)\cite{Shankar13acc} and denote the duty cycles at fixed black hole mass \mbh\ at a given redshift $z$ as $U(\mbhe,z)$. As in this work we only focus on the specific redshift of $z=0.25$, we will drop the redshift dependence in the duty cycles from here onwards. Generalising \eq\ref{eq|bias} to include the central/satellite probabilities of being active we obtain
\begin{equation}
b(\mbhe)=\frac{\left[\sum_{i=1}^{N_{\rm cen}(\mbhe)}U_{\rm cen,i}(\mbhe)(\mbhe)b_{\rm cen,i}(\mbhe)+\sum_{i=1}^{N_{\rm sat}(\mbhe)}U_{\rm sat,i}(\mbhe)b_{\rm sat,i}(\mbhe)\right]}{
\left[\sum_{i=1}^{N_{\rm cen}(\mbhe)}U_{\rm cen,i}(\mbhe)+\sum_{i=1}^{N_{\rm sat}(\mbhe)}U_{\rm sat,i}(\mbhe)\right]}\, ,
    \label{eq|biasProb}
\end{equation}
where both $U_{\rm cen}(\mbhe)$ and $U_{\rm sat}(\mbhe)$ are related to the total duty cycle of active black holes by the relation $N_{\rm AGN}(\mbhe)=U(\mbhe)N(\mbhe)=U_{\rm cen}(\mbhe)N_{\rm cen}(\mbhe)+U_{\rm sat}(\mbhe)N_{\rm sat}(\mbhe)$, with $N(\mbhe)=N_{\rm cen}(\mbhe)+N_{\rm sat}(\mbhe)$ the number of central and satellite black holes in the mass bin $\mbhe$ and $\mbhe+d\mbhe$. In the limit in which all central and satellite black holes are active or share equal probabilities to be active, i.e., $U_{\rm cen}(\mbhe)=U_{\rm sat}(\mbhe)$, then \eq\ref{eq|biasProb} reduces to the special case of \eq\ref{eq|bias}.

Throughout this work we adopt as a reference the total duty cycle $U(\mbhe)$ in \eq\ref{eq|biasProb} as derived from the continuity equation model by Shankar et al. (2013)\cite{Shankar13acc}, with constant lognormal Eddington ratio distribution. The exact normalization and shape of the total duty cycle $U(\mbhe)$ depends on the minimum luminosity threshold considered and input Eddington ratio distributions $P(\lambda)$, as anticipated in the previous \sect. As extensively discussed by, e.g., Shankar et al.\cite{Shankar13acc}, continuity equation models generally tend to generate duty cycles decreasing with increasing black hole mass, as also retrieved by direct data modelling by other groups\cite{Shankar13acc}. The latter trend would then imply that active lower mass black holes would have a larger weight on the bias (as in \figu\ref{eq|CCF} discussed below). However, when computing the mean bias in narrow bins of black hole mass (as in \figu\ref{fig|FigureBias}), the exact shape or normalization of the assumed total duty cycle $U(\mbhe)$ are irrelevant. It is in fact clear from \eq\ref{eq|biasProb} that, in a relatively small bin of black hole mass, what contributes to the mean bias is not the total duty cycle $U(\mbhe)$, rather the \emph{relative} probabilities for central and satellite black holes to be active, i.e., the ratio $Q_{\rm bh}(\mbhe)=U_{\rm sat}(\mbhe)/U_{\rm cen}(\mbhe)$.

Although satellite black holes are usually relatively less abundant than central black holes of similar mass, especially at progressively larger masses of interest to this work, if the ratio $Q_{\rm bh}\gtrsim 1$, satellite black holes could still noticeably contribute to the total mean bias as they inhabit very massive host dark matter haloes. Several works in the literature\cite{Starikova11,Shen13HOD,Lea15,Rodri17} have constrained the relative fraction of active black holes residing in satellite haloes to be around $f_{\rm sat}^{\rm AGN}\lesssim 0.1-0.2$. In terms of duty cycles, at any given black hole mass, the AGN satellite fraction $f_{\rm sat}^{\rm AGN}$ would translate into
\begin{equation}
f_{\rm sat}^{\rm AGN}(\mbhe)=\frac{U_{\rm sat}(\mbhe)N_{\rm sat}(\mbhe)}{U(\mbhe)N(\mbhe)}=
\frac{[Q_{\rm bh}(\mbhe)N_{\rm sat}(\mbhe)/N_{\rm cen}(\mbhe)]}{[Q_{\rm bh}(\mbhe)N_{\rm sat}(\mbhe)/N_{\rm cen}(\mbhe)+1]}\, ,
\label{eq|fsatAGN}
\end{equation}
implying
\begin{equation}
Q_{\rm bh}(\mbhe)=\frac{f_{\rm sat}^{\rm AGN}(\mbhe)}{1-f_{\rm sat}^{\rm AGN}(\mbhe)}\frac{1-f_{\rm sat}^{\rm BH}}{f_{\rm sat}^{\rm BH}}\, ,
\label{eq|Qbh}
\end{equation}
where $f_{\rm sat}^{\rm BH}(\mbhe)=N_{\rm sat}(\mbhe)/[N_{\rm sat}(\mbhe)+N_{\rm cen}(\mbhe)]$ is the total fraction of satellite black holes with mass within \mbh\ and $\mbhe+d\mbhe$. Based on our mocks, the total fraction of satellites for the high-mass black holes considered by Krumpe et al.\cite{Krumpe15}, is $f_{\rm sat}^{\rm BH}(\mbhe)\sim 0.1$, which corresponds to $Q_{\rm bh}(\mbhe)\sim 2$ when adopting the best-fit $f_{\rm sat}^{\rm AGN}=0.18$ from Leauthaud et al.\cite{Lea15}. We will always adopt the latter value of $Q_{\rm bh}=2$ independent of black hole mass as our reference value throughout this work, though we will also show results with $Q_{\rm bh}=1$. Indeed, $Q_{\rm bh}$ could take on even lower values. Other groups measured a $f_{\rm sat}^{\rm AGN}$ consistent with zero\cite{Starikova11}. More recently Man et al.\cite{Man19AGNSDSS} found the SDSS completeness-corrected fraction of local AGN at fixed stellar mass is independent of environment, with the fraction of AGN in satellite galaxies being comparable, if not lower, than that in central galaxies (see their \figu4). The Man et al. results would imply $Q_{\rm bh}\lesssim 1$ largely independent of host stellar mass, as assumed here.

When computing the mean bias in \figu\ref{fig|FigureBias} we make use of \eq\ref{eq|biasProb} with $Q_{\rm bh}=2$ and the halo bias $b_h$ is taken from Tinker et al. (2005)\cite{Tinker05}, in line with the Halo Occupation Distribution models adopted by Krumpe et al.\cite{Krumpe15}. We note that Tinker et al. make use of FoF-based halo masses. We thus correct our reference halo masses, defined as 200 times the critical density, by an average factor\cite{White91} of 0.966 before applying the Tinker et al. analytic model, though this has a very minor effect on the results.
For completeness, we proceed below to a more thorough comparison with the Krumpe et al.\cite{Krumpe15} full cross-correlation function between RASS (SDSS) AGN and the Luminous Red Galaxies (LRGs), as shown with filled black squares in \figu\ref{fig|CCF} for the SDSS Data Release 4 (top panels) and Data Release 7 (bottom panels). The (projected) cross-correlation function at large scales is analytically computed at any redshift of interest $z$ as
\begin{equation}
    w_P(r_P,z)=b_{\rm AGN}b_{\rm LRG}w_{\rm DM}(r_P,z) \, ,
    \label{eq|CCF}
\end{equation}
where $b_{\rm AGN}$ and $b_{\rm LRG}$ are the median large scale biases of AGN and LRGs, respectively, and $w_{\rm DM}(r_P,z)$ is the projected linear matter correlation function derived from the linear matter power spectrum of Smith et al. (2003)\cite{Smith03}, though we verified that other power spectra yield nearly identical results.

The bias of LRGs is computed by fitting the predicted linear auto-correlation function
\begin{equation}
    w_P(r_P,z)=b_{\rm LRG}^2 w_{\rm DM}(r_P,z) \, ,
    \label{eq|ACF}
\end{equation}
to the data.
Including the full covariance matrix of the uncertainty on the pair counts, a direct $\chi^2$ fitting\cite{GouldChi2} yields $b_{\rm LRG}=2.25\pm0.03$ for both DR4 and DR7. The covariance matrix was estimated in Krumpe et al.\cite{Krumpe15} (see their Equation 6) by using jackknife resampling, dividing the survey area in 100 subsection (131) for X-ray (optically) selected AGN.
The estimated value of the bias for the LRGs would imply an effective host halo mass of $M_{\rm 200c} \sim  6\times 10^{13}\, \msune/h$ and a mean
stellar mass of $\mstare\sim 5\times 10^{11}\, \msune$, in good agreement with the independent analysis by van Uitert et al. 2015\cite{Uit15} on the LOWZ and CMASS LRGs from the Baryon Oscillation Spectroscopic Survey.

Fixing $b_{\rm LRG}=2.25$ in \eq\ref{eq|CCF}, we find, when integrating over a scale of $3-30$ Mpc and including the full covariance in the $\chi^2$ fitting, $b_{\rm AGN}=1.43\pm0.11$ and $b_{\rm AGN}=1.56\pm0.10$ for DR7 and DR4, respectively. In our mocks we then select a black hole mass distribution consistent with the one adopted by Krumpe et al., i.e., with a lognormal distribution peaked at $\log \mbhe/\msune=8.5$ and with a dispersion of 0.3 dex (the ``high $M_{\rm BH}$ sample'' in their \figu5). We then compute the mean bias as given in \eq\ref{eq|biasProb} (we stress that medians would, if anything, tend to produce lower biases) assuming $Q_{\rm bh}=1$ (left panels) and $Q_{\rm bh}=2$ (right panels), which would correspond to a fraction $f_{\rm sat}\sim 0.1-0.2$, following \eq\ref{eq|Qbh}. When computing the full cross-correlation function we use the more recent analytic model for the bias put forward by Tinker et al. (2010)\cite{Tinker10}, which provides a closer match to the halo autocorrelation function from the MultiDark simulation\cite{Klypin16}, as shown in Supplementary \figu\ref{fig|BiasHaloMultiDark}. Keeping the Tinker et al. (2005) bias would, if anything, further strengthen our conclusion that the clustering data are most consistent with the intrinsic/unbiased black hole scaling relations.

\eq\ref{eq|biasProb} yields a mean $b_{\rm AGN}=1.39$ and $b_{\rm AGN}=1.08$, for $Q_{\rm bh}=1$ and $b_{\rm AGN}=1.42$ and $b_{\rm AGN}=1.13$, for $Q_{\rm bh}=2$, when adopting the intrinsic and observed \mbh-\mstar\ relations, \eqs\ref{eq|IntrinsicMbhMstar} and \ref{eq|ObservedMbhMstar}, respectively. The latter model is always $\gtrsim 3-3.5\sigma$ away from the best-fit value of the DR7 (and even more for the DR4), in line with the results presented in \figu\ref{fig|FigureBias}.
We note that all our assumptions so far have been as conservative as possible, aimed at minimizing the tension in the large-scale clustering between the data and the model based on the observed \mbh-\mstar\ relation. Adopting even lower $Q_{\rm bh}$ values, and/or more up-to-date observed \mbh-\mstar\ relations\cite{Davis18}, and/or empirically-based duty cycles\cite{Schulze15}, would all tend to yield mean values of the predicted $b_{\rm AGN}$ close to unity, implying $>3-4\sigma$ departures from the mean bias directly retrieved from the projected correlation function.
The implied cross-correlation functions from the intrinsic and observed models, found by setting $b_{\rm LRG}=2.25$ into \eq\ref{eq|CCF}, are reported in the Supplementary \figu\ref{fig|CCF} with solid red and long-dashed black lines, respectively.

\subsection{Systematics in mean radiative efficiency and obscured fractions}
\label{subsec|systematics}

This work puts forward a novel methodology that makes use of AGN clustering measurements to set firmer constraints on the shape of the local \mbh-\mstar\ relation and, in turn, on the mean radiative efficiency of AGN. Some key points should be emphasized. When comparing to a given rendition of the local \mbh-\mstar\ relation, the accretion models are actually able to constrain the \emph{ratio} between bolometric correction and radiative efficiency $\kbe(1-\epsie)/\epsie$ (see \eq\ref{eq|Soltan}).

We showed in \figu\ref{fig|FigureRhoz} that, when adopting the latest renditions of the (luminosity-dependent) bolometric corrections by Yang et al.\cite{Lusso12,Yang18}, we require mean radiative efficiencies of the order of $\epsie\gtrsim 0.1-0.2$ to match the local black hole mass density implied by the intrinsic/unbiased \mbh-\mstar\ scaling relation\cite{Shankar16BH}. Up to one order of magnitude lower mean radiative efficiencies $\epsie\sim 0.02$, would instead be necessary to broadly align with the local black hole mass density inferred from the \emph{observed} \mbh-\mstar\ relation (yellow region in the right panel of \figu\ref{fig|FigureBHMF}).

Ueda et al.\cite{Ueda14} performed very similar calculations to ours and with the same AGN luminosity function. When adopting as a local reference a black hole mass function derived from the Kormendy \& Ho \mbh-\mstar\ scaling relation, comparable to the dot-dashed black lines in \figu\ref{fig|FigureBHMF} (filled red circles in their \figu23), they derived a best-fit radiative efficiency of $\epsie=0.053$. Correcting their estimate of the mean radiative efficiency based on a different bolometric correction\cite{Hop07}, which is a factor of $\sim 2-3$ higher than our reference one, would imply extremely low values of $\epsie\lesssim 0.03$, fully consistent with our ``biased'' estimate of the radiative efficiency. The previous results by Ueda et al. and other works would thus be in tension with estimates of independent spectral estimates of the spin of supermassive black holes\cite{Reynolds19}, which suggest radiative efficiencies of the order of $\epsie \gtrsim 0.1$ or above.

Higher values of the bolometric corrections would in fact require proportionally higher radiative efficiencies to generate the same relic black hole mass density (\eq\ref{eq|Soltan}). Indeed, Zhang et al.\cite{ZhangLu19eta} recently retrieved a mean radiative efficiency of $\epsie\sim 0.3$ when adopting Shankar et al.'s intrinsic local black hole relations, which is in line with our estimates when considering their $\sim 2-3$ larger values of the radiative efficiency\cite{Hop07,ZhangLu12}.

Shankar et al. (2013)\cite{Shankar13acc} also considered alternative accretion models that made use of either an Eddington ratio-dependent bolometric correction from Vasudevan \& Fabian\cite{Vasudevan07}, or a redshift-dependent mean radiative efficiency. The former model produced a relic black hole mass function similar in shape to the one based on luminosity-dependent bolometric corrections (see their \figu14), though with noticeably less massive black holes (their \figu15), implying an overall lower integrated black hole mass density falling in between the models presented in the top and bottom panels of \figu\ref{fig|FigureRhoz}. A mean radiative efficiency decreasing with cosmic time from close to maximal ($\epsie\sim 0.2-0.3$) to minimal values ($\epsie\sim 0.05$) at low redshift, was originally considered by Shankar et al. as a possibility to reconcile the preliminary analysis of high-redshift clustering data of luminous quasars, which already suggested relatively high radiative efficiencies\cite{ShankarCrocce}, with the ``biased'' local black hole mass densities. A progressively decreasing radiative efficiency at late times is however disfavoured by our current results as it would generate relic black hole mass densities too large with respect to our updated local black hole mass densities.

Another important point to mention, which can be of relevance for all statistical supermassive black hole accretion models, is the possible presence of a selection effect on black hole spin distribution affecting flux-limited AGN surveys\cite{Vasudevan16}. Assuming an intrinsically broad spin distribution at, say, fixed black hole mass, flux-limit effects would then favour the detection of the more luminous, higher-spin/higher-\epsi\ sources. This in turn would imply that the average intrinsic radiative efficiency could be lower at fixed black hole or host galaxy stellar mass. On the other hand we note that even assuming a minimum of $\epsie=0.05$, setting\cite{SWM} $L=f_{0.1}\dot{m}l\mbhe$, with $f=\epsie/(1-\epsie)$, $l\sim 1.3\times 10^{38}\, {\rm erg\, s^{-1}\, \msune^{-1}}$, $\dot{m}=\dot{M}_{\rm bh}/\dot{M}_{\rm Edd}$, and $\dot{M}_{\rm Edd}=L_{\rm Edd}/0.1 c^2$, we would get a bolometric luminosity of $L\gtrsim 10^{43}\, \ergse$, for $\mbhe \gtrsim 10^6\, \msune$ accreting at an average\cite{Shankar13acc} $\dot{m}\sim 0.1$. This limit in bolometric luminosity would correspond to $L_X\gtrsim 10^{42}\, \ergse$, for typical X-ray bolometric corrections at these luminosities, which is still within the reach of the surveys adopted in this work, at least at $z\lesssim 2$ (see left panel of \figu3 in Ueda et al.\cite{Ueda14}).

Last but not least, a fundamental source of systematic uncertainty in the calculation of the mean radiative efficiency derived from \eq\ref{eq|Soltan}, originates from the fraction of obscured AGN included in the AGN luminosity function. \eq\ref{eq|Soltan} in fact implies specifying a \emph{complete} census of AGN, and thus the AGN luminosity function must be corrected for the fraction of obscured sources $\fobse(L,z)$ missed even in hard X-ray AGN surveys. In this work we followed Ueda et al.\cite{Ueda14}, who suggest a fraction of obscured Compton-thick AGN with column densities in the range $24<\log N_H/{\rm cm^{-2}}<26$ equal to 50\% the number densities of AGN with $20<\log N_H/{\rm cm^{-2}}<24$. The Ueda et al. estimate is in good agreement with the more recent extrapolated estimates from NuSTAR observations of heavily obscured Swift/BAT AGN by Georgantopoulos and Akylas\cite{GeoAkilas19}, who predict, from analyzing a fraction of the BAT sample, \emph{up to} 50\% of Compton-thick AGN with respect to the total population, within typical AGN spectral parameters. On the other hand, from cutting-edge comprehensive AGN population synthesis models, Ananna et al.\cite{Ana19} have recently claimed a fraction of Compton-thick AGN with $24<\log N_H/{\rm cm^{-2}}<26$ a factor of $\sim 2-3$ higher than what estimated by Ueda et al. (their \figu10). Increasing the number densities of AGN increases the normalization of the AGN luminosity function $\Phi(L,z)$ in \eq\ref{eq|Soltan}, implying a higher relic mass density at fixed bolometric correction, thus proportionally higher radiative efficiencies to match the same local black hole mass density.

We note that in hierarchical models of structure formation\cite{White91}, a substantial population of wondering black holes may end up co-existing with the central black hole within the same host galaxy. Accounting for this additional, at present unknown, population of wondering black holes in the estimate of the local black hole mass density would imply a decrease in the mean radiative efficiency if the wondering black holes were themselves active and recorded in the AGN luminosity function. More generally, orbiting and/or ejected black holes could affect the applicability of \eq\ref{eq|Soltan}. Cosmological hydrodynamic simulations\cite{Kulier15} showed, however, that the correction induced by the integrated contribution of black holes today not locked up at the centre of galaxies, should not amount to more than $\sim 11\%$, on average.


All in all, our inferred mean radiative efficiencies of $\epsie\gtrsim 0.1-0.2$ can be safely considered as lower limits at fixed kinetic efficiency $\epsie_{\rm kin}$ and local black hole mass density, as our adopted bolometric corrections and obscured fractions are comparable to, or if anything lower than, other estimates put forward in the literature.

\end{methods}




\begin{thebibliography}{10}
\expandafter\ifx\csname url\endcsname\relax
  \def\url#1{\texttt{#1}}\fi
\expandafter\ifx\csname urlprefix\endcsname\relax\def\urlprefix{URL }\fi
\providecommand{\bibinfo}[2]{#2}
\providecommand{\eprint}[2][]{\url{#2}}

\bibitem{Rees84}
\bibinfo{author}{{Rees}, M.~J.}
\newblock \bibinfo{title}{{Black Hole Models for Active Galactic Nuclei}}.
\newblock \emph{\bibinfo{journal}{\araa}} \textbf{\bibinfo{volume}{22}},
  \bibinfo{pages}{471--506} (\bibinfo{year}{1984}).

\bibitem{Bardeen72}
\bibinfo{author}{{Bardeen}, J.~M.}, \bibinfo{author}{{Press}, W.~H.} \&
  \bibinfo{author}{{Teukolsky}, S.~A.}
\newblock \bibinfo{title}{{Rotating Black Holes: Locally Nonrotating Frames,
  Energy Extraction, and Scalar Synchrotron Radiation}}.
\newblock \emph{\bibinfo{journal}{\apj}} \textbf{\bibinfo{volume}{178}},
  \bibinfo{pages}{347--370} (\bibinfo{year}{1972}).

\bibitem{Soltan}
\bibinfo{author}{{Soltan}, A.}
\newblock \bibinfo{title}{{Masses of quasars}}.
\newblock \emph{\bibinfo{journal}{\mnras}} \textbf{\bibinfo{volume}{200}},
  \bibinfo{pages}{115--122} (\bibinfo{year}{1982}).

\bibitem{Salucci99}
\bibinfo{author}{{Salucci}, P.}, \bibinfo{author}{{Szuszkiewicz}, E.},
  \bibinfo{author}{{Monaco}, P.} \& \bibinfo{author}{{Danese}, L.}
\newblock \bibinfo{title}{{Mass function of dormant black holes and the
  evolution of active galactic nuclei}}.
\newblock \emph{\bibinfo{journal}{\mnras}} \textbf{\bibinfo{volume}{307}},
  \bibinfo{pages}{637--644} (\bibinfo{year}{1999}).
\newblock \eprint{arXiv:astro-ph/9811102}.

\bibitem{Marconi04}
\bibinfo{author}{{Marconi}, A.} \emph{et~al.}
\newblock \bibinfo{title}{{Local supermassive black holes, relics of active
  galactic nuclei and the X-ray background}}.
\newblock \emph{\bibinfo{journal}{\mnras}} \textbf{\bibinfo{volume}{351}},
  \bibinfo{pages}{169--185} (\bibinfo{year}{2004}).
\newblock \eprint{arXiv:astro-ph/0311619}.

\bibitem{Shankar13acc}
\bibinfo{author}{{Shankar}, F.}, \bibinfo{author}{{Weinberg}, D.~H.} \&
  \bibinfo{author}{{Miralda-Escud{\'e}}, J.}
\newblock \bibinfo{title}{{Accretion-driven evolution of black holes: Eddington
  ratios, duty cycles and active galaxy fractions}}.
\newblock \emph{\bibinfo{journal}{\mnras}} \textbf{\bibinfo{volume}{428}},
  \bibinfo{pages}{421--446} (\bibinfo{year}{2013}).
\newblock \eprint{1111.3574}.

\bibitem{Aversa15}
\bibinfo{author}{{Aversa}, R.}, \bibinfo{author}{{Lapi}, A.},
  \bibinfo{author}{{de Zotti}, G.}, \bibinfo{author}{{Shankar}, F.} \&
  \bibinfo{author}{{Danese}, L.}
\newblock \bibinfo{title}{{Black Hole and Galaxy Coevolution from Continuity
  Equation and Abundance Matching}}.
\newblock \emph{\bibinfo{journal}{\apj}} \textbf{\bibinfo{volume}{810}},
  \bibinfo{pages}{74} (\bibinfo{year}{2015}).
\newblock \eprint{1507.07318}.

\bibitem{Shankar16BH}
\bibinfo{author}{{Shankar}, F.} \emph{et~al.}
\newblock \bibinfo{title}{{Selection bias in dynamically measured supermassive
  black hole samples: its consequences and the quest for the most fundamental
  relation}}.
\newblock \emph{\bibinfo{journal}{\mnras}} \textbf{\bibinfo{volume}{460}},
  \bibinfo{pages}{3119--3142} (\bibinfo{year}{2016}).
\newblock \eprint{1603.01276}.

\bibitem{KormendyHo}
\bibinfo{author}{{Kormendy}, J.} \& \bibinfo{author}{{Ho}, L.~C.}
\newblock \bibinfo{title}{{Coevolution (Or Not) of Supermassive Black Holes and
  Host Galaxies}}.
\newblock \emph{\bibinfo{journal}{\araa}} \textbf{\bibinfo{volume}{51}},
  \bibinfo{pages}{511--653} (\bibinfo{year}{2013}).
\newblock \eprint{1304.7762}.

\bibitem{Davis18}
\bibinfo{author}{{Davis}, B.~L.}, \bibinfo{author}{{Graham}, A.~W.} \&
  \bibinfo{author}{{Cameron}, E.}
\newblock \bibinfo{title}{{Black Hole Mass Scaling Relations for Spiral
  Galaxies. II. M $_{BH}${\ndash}M $_{*,tot}$ and M $_{BH}${\ndash}M
  $_{*,disk}$}}.
\newblock \emph{\bibinfo{journal}{\apj}} \textbf{\bibinfo{volume}{869}},
  \bibinfo{pages}{113} (\bibinfo{year}{2018}).
\newblock \eprint{1810.04888}.

\bibitem{busch2014low}
\bibinfo{author}{{Busch}, G.} \emph{et~al.}
\newblock \bibinfo{title}{{A low-luminosity type-1 QSO sample. I. Overluminous
  host spheroidals or undermassive black holes}}.
\newblock \emph{\bibinfo{journal}{\aap}} \textbf{\bibinfo{volume}{561}},
  \bibinfo{pages}{A140} (\bibinfo{year}{2014}).
\newblock \eprint{1310.0272}.

\bibitem{ReinesVolonteri15}
\bibinfo{author}{{Reines}, A.~E.} \& \bibinfo{author}{{Volonteri}, M.}
\newblock \bibinfo{title}{{Relations between Central Black Hole Mass and Total
  Galaxy Stellar Mass in the Local Universe}}.
\newblock \emph{\bibinfo{journal}{\apj}} \textbf{\bibinfo{volume}{813}},
  \bibinfo{pages}{82} (\bibinfo{year}{2015}).
\newblock \eprint{1508.06274}.

\bibitem{Shankar19BH}
\bibinfo{author}{{Shankar}, F.} \emph{et~al.}
\newblock \bibinfo{title}{{Black hole scaling relations of active and quiescent
  galaxies: Addressing selection effects and constraining virial factors}}.
\newblock \emph{\bibinfo{journal}{\mnras}}  (\bibinfo{year}{2019}).
\newblock \eprint{1901.11036}.

\bibitem{Bernardi07}
\bibinfo{author}{{Bernardi}, M.}, \bibinfo{author}{{Sheth}, R.~K.},
  \bibinfo{author}{{Tundo}, E.} \& \bibinfo{author}{{Hyde}, J.~B.}
\newblock \bibinfo{title}{{Selection Bias in the M$_{\odot}$-{$\sigma$} and
  M$_{\odot}$-L Correlations and Its Consequences}}.
\newblock \emph{\bibinfo{journal}{\apj}} \textbf{\bibinfo{volume}{660}},
  \bibinfo{pages}{267--275} (\bibinfo{year}{2007}).
\newblock \eprint{arXiv:astro-ph/0609300}.

\bibitem{DaiMbhSigma}
\bibinfo{author}{{Morabito}, L.~K.} \& \bibinfo{author}{{Dai}, X.}
\newblock \bibinfo{title}{{A Bayesian Monte Carlo Analysis of the M-{$\sigma$}
  Relation}}.
\newblock \emph{\bibinfo{journal}{\apj}} \textbf{\bibinfo{volume}{757}},
  \bibinfo{pages}{172} (\bibinfo{year}{2012}).
\newblock \eprint{1208.3900}.

\bibitem{CooraySheth}
\bibinfo{author}{{Cooray}, A.} \& \bibinfo{author}{{Sheth}, R.}
\newblock \bibinfo{title}{{Halo models of large scale structure}}.
\newblock \emph{\bibinfo{journal}{\physrep}} \textbf{\bibinfo{volume}{372}},
  \bibinfo{pages}{1--129} (\bibinfo{year}{2002}).
\newblock \eprint{arXiv:astro-ph/0206508}.

\bibitem{Shankar17g}
\bibinfo{author}{{Shankar}, F.} \emph{et~al.}
\newblock \bibinfo{title}{{Revisiting the Bulge{\ndash}Halo Conspiracy. I.
  Dependence on Galaxy Properties and Halo Mass}}.
\newblock \emph{\bibinfo{journal}{\apj}} \textbf{\bibinfo{volume}{840}},
  \bibinfo{pages}{34} (\bibinfo{year}{2017}).
\newblock \eprint{1703.06145}.

\bibitem{Grylls19}
\bibinfo{author}{{Grylls}, P.~J.}, \bibinfo{author}{{Shankar}, F.},
  \bibinfo{author}{{Zanisi}, L.} \& \bibinfo{author}{{Bernardi}, M.}
\newblock \bibinfo{title}{{A statistical semi-empirical model: satellite
  galaxies in groups and clusters}}.
\newblock \emph{\bibinfo{journal}{\mnras}} \textbf{\bibinfo{volume}{483}},
  \bibinfo{pages}{2506--2523} (\bibinfo{year}{2019}).
\newblock \eprint{1812.00015}.

\bibitem{Dave19}
\bibinfo{author}{{Dav{\'e}}, R.} \emph{et~al.}
\newblock \bibinfo{title}{{Simba: Cosmological Simulations with Black Hole
  Growth and Feedback}}.
\newblock \emph{\bibinfo{journal}{arXiv e-prints}}  (\bibinfo{year}{2019}).
\newblock \eprint{1901.10203}.

\bibitem{Savorgnan16}
\bibinfo{author}{{Savorgnan}, G.~A.~D.}, \bibinfo{author}{{Graham}, A.~W.},
  \bibinfo{author}{{Marconi}, A.} \& \bibinfo{author}{{Sani}, E.}
\newblock \bibinfo{title}{{Supermassive Black Holes and Their Host Spheroids.
  II. The Red and Blue Sequence in the M$_{BH}$-M$_{*,sph}$ Diagram}}.
\newblock \emph{\bibinfo{journal}{\apj}} \textbf{\bibinfo{volume}{817}},
  \bibinfo{pages}{21} (\bibinfo{year}{2016}).
\newblock \eprint{1511.07437}.

\bibitem{Sahu19}
\bibinfo{author}{{Sahu}, N.}, \bibinfo{author}{{Graham}, A.~W.} \&
  \bibinfo{author}{{Davis}, B.~L.}
\newblock \bibinfo{title}{{Black Hole Mass Scaling Relations for Early-Type
  Galaxies: $M\_{BH}$-$M\_{*,sph}$ and $M\_{BH}$-$M\_{*,gal}$}}.
\newblock \emph{\bibinfo{journal}{arXiv e-prints}}  (\bibinfo{year}{2019}).
\newblock \eprint{1903.04738}.

\bibitem{Baron19}
\bibinfo{author}{{Baron}, D.} \& \bibinfo{author}{{M{\'e}nard}, B.}
\newblock \bibinfo{title}{{Black hole mass estimation for Active Galactic
  Nuclei from a new angle}}.
\newblock \emph{\bibinfo{journal}{arXiv e-prints}}
  \bibinfo{pages}{arXiv:1903.01996} (\bibinfo{year}{2019}).
\newblock \eprint{1903.01996}.

\bibitem{Powell18}
\bibinfo{author}{{Powell}, M.~C.} \emph{et~al.}
\newblock \bibinfo{title}{{The Swift/BAT AGN Spectroscopic Survey. IX. The
  Clustering Environments of an Unbiased Sample of Local AGNs}}.
\newblock \emph{\bibinfo{journal}{\apj}} \textbf{\bibinfo{volume}{858}},
  \bibinfo{pages}{110} (\bibinfo{year}{2018}).
\newblock \eprint{1803.07589}.

\bibitem{Krumpe15}
\bibinfo{author}{{Krumpe}, M.} \emph{et~al.}
\newblock \bibinfo{title}{{The Spatial Clustering of ROSAT All-Sky Survey
  Active Galactic Nuclei. IV. More Massive Black Holes Reside in More Massive
  Dark Matter Halos}}.
\newblock \emph{\bibinfo{journal}{\apj}} \textbf{\bibinfo{volume}{815}},
  \bibinfo{pages}{21} (\bibinfo{year}{2015}).
\newblock \eprint{1509.01261}.

\bibitem{Krumpe18}
\bibinfo{author}{{Krumpe}, M.}, \bibinfo{author}{{Miyaji}, T.},
  \bibinfo{author}{{Coil}, A.~L.} \& \bibinfo{author}{{Aceves}, H.}
\newblock \bibinfo{title}{{Spatial clustering and halo occupation distribution
  modelling of local AGN via cross-correlation measurements with 2MASS
  galaxies}}.
\newblock \emph{\bibinfo{journal}{\mnras}} \textbf{\bibinfo{volume}{474}},
  \bibinfo{pages}{1773--1786} (\bibinfo{year}{2018}).
\newblock \eprint{1710.05638}.

\bibitem{Ghez08}
\bibinfo{author}{{Ghez}, A.~M.} \emph{et~al.}
\newblock \bibinfo{title}{{Measuring Distance and Properties of the Milky Way's
  Central Supermassive Black Hole with Stellar Orbits}}.
\newblock \emph{\bibinfo{journal}{\apj}} \textbf{\bibinfo{volume}{689}},
  \bibinfo{pages}{1044--1062} (\bibinfo{year}{2008}).
\newblock \eprint{0808.2870}.

\bibitem{Posti19MW}
\bibinfo{author}{{Posti}, L.} \& \bibinfo{author}{{Helmi}, A.}
\newblock \bibinfo{title}{{Mass and shape of the Milky Way's dark matter halo
  with globular clusters from Gaia and Hubble}}.
\newblock \emph{\bibinfo{journal}{\aap}} \textbf{\bibinfo{volume}{621}},
  \bibinfo{pages}{A56} (\bibinfo{year}{2019}).
\newblock \eprint{1805.01408}.

\bibitem{Shankar17BH}
\bibinfo{author}{{Shankar}, F.}, \bibinfo{author}{{Bernardi}, M.} \&
  \bibinfo{author}{{Sheth}, R.~K.}
\newblock \bibinfo{title}{{Selection bias in dynamically-measured super-massive
  black hole samples: dynamical masses and dependence on S{\'e}rsic index}}.
\newblock \emph{\bibinfo{journal}{\mnras}}  (\bibinfo{year}{2017}).
\newblock \eprint{1701.01732}.

\bibitem{Sarria10}
\bibinfo{author}{{Sarria}, J.~E.} \emph{et~al.}
\newblock \bibinfo{title}{{The M$_{BH}$ - M$_{star}$ relation of obscured AGNs
  at high redshift}}.
\newblock \emph{\bibinfo{journal}{\aap}} \textbf{\bibinfo{volume}{522}},
  \bibinfo{pages}{L3} (\bibinfo{year}{2010}).
\newblock \eprint{1010.0768}.

\bibitem{Falomo14}
\bibinfo{author}{{Falomo}, R.}, \bibinfo{author}{{Bettoni}, D.},
  \bibinfo{author}{{Karhunen}, K.}, \bibinfo{author}{{Kotilainen}, J.~K.} \&
  \bibinfo{author}{{Uslenghi}, M.}
\newblock \bibinfo{title}{{Low-redshift quasars in the Sloan Digital Sky Survey
  Stripe 82. The host galaxies}}.
\newblock \emph{\bibinfo{journal}{\mnras}} \textbf{\bibinfo{volume}{440}},
  \bibinfo{pages}{476--493} (\bibinfo{year}{2014}).
\newblock \eprint{1402.4300}.

\bibitem{Tinker08}
\bibinfo{author}{{Tinker}, J.} \emph{et~al.}
\newblock \bibinfo{title}{{Toward a Halo Mass Function for Precision Cosmology:
  The Limits of Universality}}.
\newblock \emph{\bibinfo{journal}{\apj}} \textbf{\bibinfo{volume}{688}},
  \bibinfo{pages}{709--728} (\bibinfo{year}{2008}).
\newblock \eprint{0803.2706}.

\bibitem{DiPompeo16}
\bibinfo{author}{{DiPompeo}, M.~A.}, \bibinfo{author}{{Runnoe}, J.~C.},
  \bibinfo{author}{{Hickox}, R.~C.}, \bibinfo{author}{{Myers}, A.~D.} \&
  \bibinfo{author}{{Geach}, J.~E.}
\newblock \bibinfo{title}{{The impact of the dusty torus on obscured quasar
  halo mass measurements}}.
\newblock \emph{\bibinfo{journal}{\mnras}} \textbf{\bibinfo{volume}{460}},
  \bibinfo{pages}{175--186} (\bibinfo{year}{2016}).
\newblock \eprint{1604.06811}.

\bibitem{Jiang16bias}
\bibinfo{author}{{Jiang}, N.} \emph{et~al.}
\newblock \bibinfo{title}{{Differences in Halo-scale Environments between Type
  1 and Type 2 AGNs at Low Redshift}}.
\newblock \emph{\bibinfo{journal}{\apj}} \textbf{\bibinfo{volume}{832}},
  \bibinfo{pages}{111} (\bibinfo{year}{2016}).
\newblock \eprint{1602.08825}.

\bibitem{Shankar10shen}
\bibinfo{author}{{Shankar}, F.}, \bibinfo{author}{{Weinberg}, D.~H.} \&
  \bibinfo{author}{{Shen}, Y.}
\newblock \bibinfo{title}{{Constraints on black hole duty cycles and the black
  hole-halo relation from SDSS quasar clustering}}.
\newblock \emph{\bibinfo{journal}{\mnras}} \textbf{\bibinfo{volume}{406}},
  \bibinfo{pages}{1959--1966} (\bibinfo{year}{2010}).
\newblock \eprint{1004.1173}.

\bibitem{ShethTormen}
\bibinfo{author}{{Sheth}, R.~K.} \& \bibinfo{author}{{Tormen}, G.}
\newblock \bibinfo{title}{{Large-scale bias and the peak background split}}.
\newblock \emph{\bibinfo{journal}{\mnras}} \textbf{\bibinfo{volume}{308}},
  \bibinfo{pages}{119--126} (\bibinfo{year}{1999}).
\newblock \eprint{arXiv:astro-ph/9901122}.

\bibitem{Ueda14}
\bibinfo{author}{{Ueda}, Y.}, \bibinfo{author}{{Akiyama}, M.},
  \bibinfo{author}{{Hasinger}, G.}, \bibinfo{author}{{Miyaji}, T.} \&
  \bibinfo{author}{{Watson}, M.~G.}
\newblock \bibinfo{title}{{Toward the Standard Population Synthesis Model of
  the X-Ray Background: Evolution of X-Ray Luminosity and Absorption Functions
  of Active Galactic Nuclei Including Compton-thick Populations}}.
\newblock \emph{\bibinfo{journal}{\apj}} \textbf{\bibinfo{volume}{786}},
  \bibinfo{pages}{104} (\bibinfo{year}{2014}).
\newblock \eprint{1402.1836}.

\bibitem{Yang18}
\bibinfo{author}{{Yang}, G.} \emph{et~al.}
\newblock \bibinfo{title}{{Linking black hole growth with host galaxies: the
  accretion-stellar mass relation and its cosmic evolution}}.
\newblock \emph{\bibinfo{journal}{\mnras}} \textbf{\bibinfo{volume}{475}},
  \bibinfo{pages}{1887--1911} (\bibinfo{year}{2018}).
\newblock \eprint{1710.09399}.

\bibitem{Harrison16}
\bibinfo{author}{{Harrison}, F.~A.} \emph{et~al.}
\newblock \bibinfo{title}{{The NuSTAR Extragalactic Surveys: The Number Counts
  of Active Galactic Nuclei and the Resolved Fraction of the Cosmic X-Ray
  Background}}.
\newblock \emph{\bibinfo{journal}{\apj}} \textbf{\bibinfo{volume}{831}},
  \bibinfo{pages}{185} (\bibinfo{year}{2016}).
\newblock \eprint{1511.04183}.

\bibitem{Shankar08Cav}
\bibinfo{author}{{Shankar}, F.}, \bibinfo{author}{{Cavaliere}, A.},
  \bibinfo{author}{{Cirasuolo}, M.} \& \bibinfo{author}{{Maraschi}, L.}
\newblock \bibinfo{title}{{Optical-Radio Mapping: the Kinetic Efficiency of
  Radio-Loud AGNs}}.
\newblock \emph{\bibinfo{journal}{\apj}} \textbf{\bibinfo{volume}{676}},
  \bibinfo{pages}{131--136} (\bibinfo{year}{2008}).
\newblock \eprint{0712.3004}.

\bibitem{Reynolds19}
\bibinfo{author}{{Reynolds}, C.~S.}
\newblock \bibinfo{title}{{Observing black holes spin}}.
\newblock \emph{\bibinfo{journal}{Nature Astronomy}}
  \textbf{\bibinfo{volume}{3}}, \bibinfo{pages}{41--47} (\bibinfo{year}{2019}).
\newblock \eprint{1903.11704}.

\bibitem{Shankar16}
\bibinfo{author}{{Shankar}, F.} \emph{et~al.}
\newblock \bibinfo{title}{{The Optical{\ndash}UV Emissivity of Quasars:
  Dependence on Black Hole Mass and Radio Loudness}}.
\newblock \emph{\bibinfo{journal}{\apjl}} \textbf{\bibinfo{volume}{818}},
  \bibinfo{pages}{L1} (\bibinfo{year}{2016}).
\newblock \eprint{1601.02021}.

\bibitem{ZhangLu19}
\bibinfo{author}{{Zhang}, X.} \& \bibinfo{author}{{Lu}, Y.}
\newblock \bibinfo{title}{{On Constraining the Growth History of Massive Black
  Holes via Their Distribution on the Spin-Mass Plane}}.
\newblock \emph{\bibinfo{journal}{arXiv e-prints}}
  \bibinfo{pages}{arXiv:1902.07056} (\bibinfo{year}{2019}).
\newblock \eprint{1902.07056}.

\bibitem{Elvis02}
\bibinfo{author}{{Elvis}, M.}, \bibinfo{author}{{Risaliti}, G.} \&
  \bibinfo{author}{{Zamorani}, G.}
\newblock \bibinfo{title}{{Most Supermassive Black Holes Must Be Rapidly
  Rotating}}.
\newblock \emph{\bibinfo{journal}{\apjl}} \textbf{\bibinfo{volume}{565}},
  \bibinfo{pages}{L75--L77} (\bibinfo{year}{2002}).
\newblock \eprint{astro-ph/0112413}.

\bibitem{YuLu08}
\bibinfo{author}{{Yu}, Q.} \& \bibinfo{author}{{Lu}, Y.}
\newblock \bibinfo{title}{{Toward Precise Constraints on the Growth of Massive
  Black Holes}}.
\newblock \emph{\bibinfo{journal}{\apj}} \textbf{\bibinfo{volume}{689}},
  \bibinfo{pages}{732--754} (\bibinfo{year}{2008}).
\newblock \eprint{0808.3777}.

\bibitem{Merloni_eRosita}
\bibinfo{author}{{Merloni}, A.} \emph{et~al.}
\newblock \bibinfo{title}{{eROSITA Science Book: Mapping the Structure of the
  Energetic Universe}}.
\newblock \emph{\bibinfo{journal}{ArXiv e-prints}}  (\bibinfo{year}{2012}).
\newblock \eprint{1209.3114}.

\end{thebibliography}

\begin{thebibliography}{10}
\expandafter\ifx\csname url\endcsname\relax
  \def\url#1{\texttt{#1}}\fi
\expandafter\ifx\csname urlprefix\endcsname\relax\def\urlprefix{URL }\fi
\providecommand{\bibinfo}[2]{#2}
\providecommand{\eprint}[2][]{\url{#2}}

\bibitem{Bell03SEDs}
\bibinfo{author}{{Bell}, E.~F.}, \bibinfo{author}{{McIntosh}, D.~H.},
  \bibinfo{author}{{Katz}, N.} \& \bibinfo{author}{{Weinberg}, M.~D.}
\newblock \bibinfo{title}{{The Optical and Near-Infrared Properties of
  Galaxies. I. Luminosity and Stellar Mass Functions}}.
\newblock \emph{\bibinfo{journal}{\apjs}} \textbf{\bibinfo{volume}{149}},
  \bibinfo{pages}{289--312} (\bibinfo{year}{2003}).
\newblock \eprint{astro-ph/0302543}.

\bibitem{Chabrier03}
\bibinfo{author}{{Chabrier}, G.}
\newblock \bibinfo{title}{{Galactic Stellar and Substellar Initial Mass
  Function}}.
\newblock \emph{\bibinfo{journal}{\pasp}} \textbf{\bibinfo{volume}{115}},
  \bibinfo{pages}{763--795} (\bibinfo{year}{2003}).
\newblock \eprint{arXiv:astro-ph/0304382}.

\bibitem{Bernardi17}
\bibinfo{author}{{Bernardi}, M.} \emph{et~al.}
\newblock \bibinfo{title}{{The high mass end of the stellar mass function:
  Dependence on stellar population models and agreement between fits to the
  light profile}}.
\newblock \emph{\bibinfo{journal}{\mnras}}  (\bibinfo{year}{2017}).
\newblock \eprint{1604.01036}.

\bibitem{Sesana16}
\bibinfo{author}{{Sesana}, A.}, \bibinfo{author}{{Shankar}, F.},
  \bibinfo{author}{{Bernardi}, M.} \& \bibinfo{author}{{Sheth}, R.~K.}
\newblock \bibinfo{title}{{Selection bias in dynamically measured supermassive
  black hole samples: consequences for pulsar timing arrays}}.
\newblock \emph{\bibinfo{journal}{\mnras}} \textbf{\bibinfo{volume}{463}},
  \bibinfo{pages}{L6--L11} (\bibinfo{year}{2016}).
\newblock \eprint{1603.09348}.

\bibitem{Shankar18}
\bibinfo{author}{{Shankar}, F.} \emph{et~al.}
\newblock \bibinfo{title}{{Revisiting the bulge-halo conspiracy - II. Towards
  explaining its puzzling dependence on redshift}}.
\newblock \emph{\bibinfo{journal}{\mnras}} \textbf{\bibinfo{volume}{475}},
  \bibinfo{pages}{2878--2890} (\bibinfo{year}{2018}).
\newblock \eprint{1711.07986}.

\bibitem{Jiang16SubMF}
\bibinfo{author}{{Jiang}, F.} \& \bibinfo{author}{{van den Bosch}, F.~C.}
\newblock \bibinfo{title}{{Statistics of dark matter substructure - I. Model
  and universal fitting functions}}.
\newblock \emph{\bibinfo{journal}{\mnras}} \textbf{\bibinfo{volume}{458}},
  \bibinfo{pages}{2848--2869} (\bibinfo{year}{2016}).

\bibitem{Giocoli08}
\bibinfo{author}{{Giocoli}, C.}, \bibinfo{author}{{Tormen}, G.} \&
  \bibinfo{author}{{van den Bosch}, F.~C.}
\newblock \bibinfo{title}{{The population of dark matter subhaloes: mass
  functions and average mass-loss rates}}.
\newblock \emph{\bibinfo{journal}{\mnras}} \textbf{\bibinfo{volume}{386}},
  \bibinfo{pages}{2135--2144} (\bibinfo{year}{2008}).
\newblock \eprint{0712.1563}.

\bibitem{Bernardi13}
\bibinfo{author}{{Bernardi}, M.} \emph{et~al.}
\newblock \bibinfo{title}{{The massive end of the luminosity and stellar mass
  functions: dependence on the fit to the light profile}}.
\newblock \emph{\bibinfo{journal}{\mnras}} \textbf{\bibinfo{volume}{436}},
  \bibinfo{pages}{697--704} (\bibinfo{year}{2013}).
\newblock \eprint{1304.7778}.

\bibitem{Bernardi16}
\bibinfo{author}{{Bernardi}, M.} \emph{et~al.}
\newblock \bibinfo{title}{{The massive end of the luminosity and stellar mass
  functions and clustering from CMASS to SDSS: evidence for and against passive
  evolution}}.
\newblock \emph{\bibinfo{journal}{\mnras}} \textbf{\bibinfo{volume}{455}},
  \bibinfo{pages}{4122--4135} (\bibinfo{year}{2016}).
\newblock \eprint{1510.07702}.

\bibitem{Tinker17}
\bibinfo{author}{{Tinker}, J.~L.} \emph{et~al.}
\newblock \bibinfo{title}{{The Correlation between Halo Mass and Stellar Mass
  for the Most Massive Galaxies in the Universe}}.
\newblock \emph{\bibinfo{journal}{\apj}} \textbf{\bibinfo{volume}{839}},
  \bibinfo{pages}{121} (\bibinfo{year}{2017}).
\newblock \eprint{1607.04678}.

\bibitem{Kravtsov18}
\bibinfo{author}{{Kravtsov}, A.~V.}, \bibinfo{author}{{Vikhlinin}, A.~A.} \&
  \bibinfo{author}{{Meshcheryakov}, A.~V.}
\newblock \bibinfo{title}{{Stellar Mass-Halo Mass Relation and Star Formation
  Efficiency in High-Mass Halos}}.
\newblock \emph{\bibinfo{journal}{Astronomy Letters}}
  \textbf{\bibinfo{volume}{44}}, \bibinfo{pages}{8--34} (\bibinfo{year}{2018}).
\newblock \eprint{1401.7329}.

\bibitem{Behroozi18}
\bibinfo{author}{{Behroozi}, P.}, \bibinfo{author}{{Wechsler}, R.},
  \bibinfo{author}{{Hearin}, A.} \& \bibinfo{author}{{Conroy}, C.}
\newblock \bibinfo{title}{{UniverseMachine: The Correlation between Galaxy
  Growth and Dark Matter Halo Assembly from z=0-10}}.
\newblock \emph{\bibinfo{journal}{ArXiv e-prints}}  (\bibinfo{year}{2018}).
\newblock \eprint{1806.07893}.

\bibitem{Moster18}
\bibinfo{author}{{Moster}, B.~P.}, \bibinfo{author}{{Naab}, T.} \&
  \bibinfo{author}{{White}, S.~D.~M.}
\newblock \bibinfo{title}{{EMERGE - an empirical model for the formation of
  galaxies since z ~ 10}}.
\newblock \emph{\bibinfo{journal}{\mnras}} \textbf{\bibinfo{volume}{477}},
  \bibinfo{pages}{1822--1852} (\bibinfo{year}{2018}).
\newblock \eprint{1705.05373}.

\bibitem{Huertas11}
\bibinfo{author}{{Huertas-Company}, M.}, \bibinfo{author}{{Aguerri}, J.~A.~L.},
  \bibinfo{author}{{Bernardi}, M.}, \bibinfo{author}{{Mei}, S.} \&
  \bibinfo{author}{{S{\'a}nchez Almeida}, J.}
\newblock \bibinfo{title}{{Revisiting the Hubble sequence in the SDSS DR7
  spectroscopic sample: a publicly available Bayesian automated
  classification}}.
\newblock \emph{\bibinfo{journal}{\aap}} \textbf{\bibinfo{volume}{525}},
  \bibinfo{pages}{A157} (\bibinfo{year}{2011}).
\newblock \eprint{1010.3018}.

\bibitem{SmallBlandford}
\bibinfo{author}{{Small}, T.~A.} \& \bibinfo{author}{{Blandford}, R.~D.}
\newblock \bibinfo{title}{{Quasar evolution and the growth of black holes}}.
\newblock \emph{\bibinfo{journal}{\mnras}} \textbf{\bibinfo{volume}{259}},
  \bibinfo{pages}{725--737} (\bibinfo{year}{1992}).

\bibitem{YuTremaine}
\bibinfo{author}{{Yu}, Q.} \& \bibinfo{author}{{Tremaine}, S.}
\newblock \bibinfo{title}{{Observational constraints on growth of massive black
  holes}}.
\newblock \emph{\bibinfo{journal}{\mnras}} \textbf{\bibinfo{volume}{335}},
  \bibinfo{pages}{965--976} (\bibinfo{year}{2002}).
\newblock \eprint{astro-ph/0203082}.

\bibitem{Shankar04}
\bibinfo{author}{{Shankar}, F.}, \bibinfo{author}{{Salucci}, P.},
  \bibinfo{author}{{Granato}, G.~L.}, \bibinfo{author}{{De Zotti}, G.} \&
  \bibinfo{author}{{Danese}, L.}
\newblock \bibinfo{title}{{Supermassive black hole demography: the match
  between the local and accreted mass functions}}.
\newblock \emph{\bibinfo{journal}{\mnras}} \textbf{\bibinfo{volume}{354}},
  \bibinfo{pages}{1020--1030} (\bibinfo{year}{2004}).
\newblock \eprint{arXiv:astro-ph/0405585}.

\bibitem{Cao10}
\bibinfo{author}{{Cao}, X.}
\newblock \bibinfo{title}{{Cosmological Evolution of Massive Black Holes:
  Effects of Eddington Ratio Distribution and Quasar Lifetime}}.
\newblock \emph{\bibinfo{journal}{\apj}} \textbf{\bibinfo{volume}{725}},
  \bibinfo{pages}{388--393} (\bibinfo{year}{2010}).
\newblock \eprint{1010.0046}.

\bibitem{YuLu04}
\bibinfo{author}{{Yu}, Q.} \& \bibinfo{author}{{Lu}, Y.}
\newblock \bibinfo{title}{{Constraints on QSO Models from a Relation between
  the QSO Luminosity Function and the Local Black Hole Mass Function}}.
\newblock \emph{\bibinfo{journal}{\apj}} \textbf{\bibinfo{volume}{602}},
  \bibinfo{pages}{603--624} (\bibinfo{year}{2004}).
\newblock \eprint{arXiv:astro-ph/0311404}.

\bibitem{Goulding10}
\bibinfo{author}{{Goulding}, A.~D.}, \bibinfo{author}{{Alexander}, D.~M.},
  \bibinfo{author}{{Lehmer}, B.~D.} \& \bibinfo{author}{{Mullaney}, J.~R.}
\newblock \bibinfo{title}{{Towards a complete census of active galactic nuclei
  in nearby galaxies: the incidence of growing black holes}}.
\newblock \emph{\bibinfo{journal}{\mnras}} \textbf{\bibinfo{volume}{406}},
  \bibinfo{pages}{597--611} (\bibinfo{year}{2010}).
\newblock \eprint{1003.3015}.

\bibitem{Shankar13review}
\bibinfo{author}{{Shankar}, F.}
\newblock \bibinfo{title}{{Black hole demography: from scaling relations to
  models}}.
\newblock \emph{\bibinfo{journal}{Classical and Quantum Gravity}}
  \textbf{\bibinfo{volume}{30}}, \bibinfo{pages}{244001}
  (\bibinfo{year}{2013}).
\newblock \eprint{1307.3289}.

\bibitem{Ghisellini13}
\bibinfo{author}{{Ghisellini}, G.}, \bibinfo{author}{{Haardt}, F.},
  \bibinfo{author}{{Della Ceca}, R.}, \bibinfo{author}{{Volonteri}, M.} \&
  \bibinfo{author}{{Sbarrato}, T.}
\newblock \bibinfo{title}{{The role of relativistic jets in the heaviest and
  most active supermassive black holes at high redshift}}.
\newblock \emph{\bibinfo{journal}{\mnras}} \textbf{\bibinfo{volume}{432}},
  \bibinfo{pages}{2818--2823} (\bibinfo{year}{2013}).
\newblock \eprint{1304.1152}.

\bibitem{Zubo18}
\bibinfo{author}{{Zubovas}, K.}
\newblock \bibinfo{title}{{AGN must be very efficient at powering outflows}}.
\newblock \emph{\bibinfo{journal}{\mnras}} \textbf{\bibinfo{volume}{479}},
  \bibinfo{pages}{3189--3196} (\bibinfo{year}{2018}).
\newblock \eprint{1806.08914}.

\bibitem{Starikova11}
\bibinfo{author}{{Starikova}, S.} \emph{et~al.}
\newblock \bibinfo{title}{{Constraining Halo Occupation Properties of X-Ray
  Active Galactic Nuclei Using Clustering of Chandra Sources in the Bo{\"o}tes
  Survey Region}}.
\newblock \emph{\bibinfo{journal}{\apj}} \textbf{\bibinfo{volume}{741}},
  \bibinfo{pages}{15} (\bibinfo{year}{2011}).
\newblock \eprint{1010.1577}.

\bibitem{Shen13HOD}
\bibinfo{author}{{Shen}, Y.} \emph{et~al.}
\newblock \bibinfo{title}{{Cross-correlation of SDSS DR7 Quasars and DR10 BOSS
  Galaxies: The Weak Luminosity Dependence of Quasar Clustering at z \~{}
  0.5}}.
\newblock \emph{\bibinfo{journal}{\apj}} \textbf{\bibinfo{volume}{778}},
  \bibinfo{pages}{98} (\bibinfo{year}{2013}).
\newblock \eprint{1212.4526}.

\bibitem{Lea15}
\bibinfo{author}{{Leauthaud}, A.} \emph{et~al.}
\newblock \bibinfo{title}{{The dark matter haloes of moderate luminosity X-ray
  AGN as determined from weak gravitational lensing and host stellar masses}}.
\newblock \emph{\bibinfo{journal}{\mnras}} \textbf{\bibinfo{volume}{446}},
  \bibinfo{pages}{1874--1888} (\bibinfo{year}{2015}).
\newblock \eprint{1410.5817}.

\bibitem{Rodri17}
\bibinfo{author}{{Rodr{\'{\i}}guez-Torres}, S.~A.} \emph{et~al.}
\newblock \bibinfo{title}{{Clustering of quasars in the first year of the
  SDSS-IV eBOSS survey: interpretation and halo occupation distribution}}.
\newblock \emph{\bibinfo{journal}{\mnras}} \textbf{\bibinfo{volume}{468}},
  \bibinfo{pages}{728--740} (\bibinfo{year}{2017}).
\newblock \eprint{1612.06918}.

\bibitem{Man19AGNSDSS}
\bibinfo{author}{{Man}, Z.-y.} \emph{et~al.}
\newblock \bibinfo{title}{{The Dependence of AGN Activity on Environment in
  SDSS}}.
\newblock \emph{\bibinfo{journal}{\mnras}} \bibinfo{pages}{1665}
  (\bibinfo{year}{2019}).
\newblock \eprint{1907.01563}.

\bibitem{Tinker05}
\bibinfo{author}{{Tinker}, J.~L.}, \bibinfo{author}{{Weinberg}, D.~H.},
  \bibinfo{author}{{Zheng}, Z.} \& \bibinfo{author}{{Zehavi}, I.}
\newblock \bibinfo{title}{{On the Mass-to-Light Ratio of Large-Scale
  Structure}}.
\newblock \emph{\bibinfo{journal}{\apj}} \textbf{\bibinfo{volume}{631}},
  \bibinfo{pages}{41--58} (\bibinfo{year}{2005}).
\newblock \eprint{arXiv:astro-ph/0411777}.

\bibitem{White91}
\bibinfo{author}{{White}, S.~D.~M.} \& \bibinfo{author}{{Frenk}, C.~S.}
\newblock \bibinfo{title}{{Galaxy formation through hierarchical clustering}}.
\newblock \emph{\bibinfo{journal}{\apj}} \textbf{\bibinfo{volume}{379}},
  \bibinfo{pages}{52--79} (\bibinfo{year}{1991}).

\bibitem{Smith03}
\bibinfo{author}{{Smith}, R.~E.} \emph{et~al.}
\newblock \bibinfo{title}{{Stable clustering, the halo model and non-linear
  cosmological power spectra}}.
\newblock \emph{\bibinfo{journal}{\mnras}} \textbf{\bibinfo{volume}{341}},
  \bibinfo{pages}{1311--1332} (\bibinfo{year}{2003}).
\newblock \eprint{astro-ph/0207664}.

\bibitem{GouldChi2}
\bibinfo{author}{{Gould}, A.}
\newblock \bibinfo{title}{{chi\^2 and Linear Fits}}.
\newblock \emph{\bibinfo{journal}{arXiv e-prints}}
  \bibinfo{pages}{astro--ph/0310577} (\bibinfo{year}{2003}).
\newblock \eprint{astro-ph/0310577}.

\bibitem{Uit15}
\bibinfo{author}{{van Uitert}, E.}, \bibinfo{author}{{Cacciato}, M.},
  \bibinfo{author}{{Hoekstra}, H.} \& \bibinfo{author}{{Herbonnet}, R.}
\newblock \bibinfo{title}{{Evolution of the luminosity-to-halo mass relation of
  LRGs from a combined analysis of SDSS-DR10+RCS2}}.
\newblock \emph{\bibinfo{journal}{\aap}} \textbf{\bibinfo{volume}{579}},
  \bibinfo{pages}{A26} (\bibinfo{year}{2015}).
\newblock \eprint{1503.08647}.

\bibitem{Tinker10}
\bibinfo{author}{{Tinker}, J.~L.} \emph{et~al.}
\newblock \bibinfo{title}{{The Large-scale Bias of Dark Matter Halos: Numerical
  Calibration and Model Tests}}.
\newblock \emph{\bibinfo{journal}{\apj}} \textbf{\bibinfo{volume}{724}},
  \bibinfo{pages}{878--886} (\bibinfo{year}{2010}).
\newblock \eprint{1001.3162}.

\bibitem{Klypin16}
\bibinfo{author}{{Klypin}, A.}, \bibinfo{author}{{Yepes}, G.},
  \bibinfo{author}{{Gottl{\"o}ber}, S.}, \bibinfo{author}{{Prada}, F.} \&
  \bibinfo{author}{{He{\ss}}, S.}
\newblock \bibinfo{title}{{MultiDark simulations: the story of dark matter halo
  concentrations and density profiles}}.
\newblock \emph{\bibinfo{journal}{\mnras}} \textbf{\bibinfo{volume}{457}},
  \bibinfo{pages}{4340--4359} (\bibinfo{year}{2016}).
\newblock \eprint{1411.4001}.

\bibitem{Schulze15}
\bibinfo{author}{{Schulze}, A.} \emph{et~al.}
\newblock \bibinfo{title}{{The cosmic growth of the active black hole
  population at 1<z<2 in zCOSMOS, VVDS and SDSS}}.
\newblock \emph{\bibinfo{journal}{\mnras}} \textbf{\bibinfo{volume}{447}},
  \bibinfo{pages}{2085--2111} (\bibinfo{year}{2015}).
\newblock \eprint{1412.0754}.

\bibitem{Lusso12}
\bibinfo{author}{{Lusso}, E.} \emph{et~al.}
\newblock \bibinfo{title}{{Bolometric luminosities and Eddington ratios of
  X-ray selected active galactic nuclei in the XMM-COSMOS survey}}.
\newblock \emph{\bibinfo{journal}{\mnras}} \textbf{\bibinfo{volume}{425}},
  \bibinfo{pages}{623--640} (\bibinfo{year}{2012}).
\newblock \eprint{1206.2642}.

\bibitem{Hop07}
\bibinfo{author}{{Hopkins}, P.~F.}, \bibinfo{author}{{Richards}, G.~T.} \&
  \bibinfo{author}{{Hernquist}, L.}
\newblock \bibinfo{title}{{An Observational Determination of the Bolometric
  Quasar Luminosity Function}}.
\newblock \emph{\bibinfo{journal}{\apj}} \textbf{\bibinfo{volume}{654}},
  \bibinfo{pages}{731--753} (\bibinfo{year}{2007}).
\newblock \eprint{arXiv:astro-ph/0605678}.

\bibitem{ZhangLu19eta}
\bibinfo{author}{{Zhang}, X.} \& \bibinfo{author}{{Lu}, Y.}
\newblock \bibinfo{title}{{On the mean radiative efficiency of accreting
  massive black holes in AGNs and QSOs}}.
\newblock \emph{\bibinfo{journal}{Science China Physics, Mechanics, and
  Astronomy}} \textbf{\bibinfo{volume}{60}}, \bibinfo{pages}{109511}
  (\bibinfo{year}{2017}).
\newblock \eprint{1902.08332}.

\bibitem{ZhangLu12}
\bibinfo{author}{{Zhang}, X.}, \bibinfo{author}{{Lu}, Y.} \&
  \bibinfo{author}{{Yu}, Q.}
\newblock \bibinfo{title}{{The Cosmic Evolution of Massive Black Holes and
  Galaxy Spheroids: Global Constraints at Redshift z < 1.2}}.
\newblock \emph{\bibinfo{journal}{\apj}} \textbf{\bibinfo{volume}{761}},
  \bibinfo{pages}{5} (\bibinfo{year}{2012}).
\newblock \eprint{1210.4019}.

\bibitem{Vasudevan07}
\bibinfo{author}{{Vasudevan}, R.~V.} \& \bibinfo{author}{{Fabian}, A.~C.}
\newblock \bibinfo{title}{{Piecing together the X-ray background: bolometric
  corrections for active galactic nuclei}}.
\newblock \emph{\bibinfo{journal}{\mnras}} \textbf{\bibinfo{volume}{381}},
  \bibinfo{pages}{1235--1251} (\bibinfo{year}{2007}).
\newblock \eprint{0708.4308}.

\bibitem{ShankarCrocce}
\bibinfo{author}{{Shankar}, F.}, \bibinfo{author}{{Crocce}, M.},
  \bibinfo{author}{{Miralda-Escud{\'e}}, J.}, \bibinfo{author}{{Fosalba}, P.}
  \& \bibinfo{author}{{Weinberg}, D.~H.}
\newblock \bibinfo{title}{{On the Radiative Efficiencies, Eddington Ratios, and
  Duty Cycles of Luminous High-redshift Quasars}}.
\newblock \emph{\bibinfo{journal}{\apj}} \textbf{\bibinfo{volume}{718}},
  \bibinfo{pages}{231--250} (\bibinfo{year}{2010}).
\newblock \eprint{0810.4919}.

\bibitem{Vasudevan16}
\bibinfo{author}{{Vasudevan}, R.~V.} \emph{et~al.}
\newblock \bibinfo{title}{{A selection effect boosting the contribution from
  rapidly spinning black holes to the cosmic X-ray background}}.
\newblock \emph{\bibinfo{journal}{\mnras}} \textbf{\bibinfo{volume}{458}},
  \bibinfo{pages}{2012--2023} (\bibinfo{year}{2016}).
\newblock \eprint{1506.01027}.

\bibitem{SWM}
\bibinfo{author}{{Shankar}, F.}, \bibinfo{author}{{Weinberg}, D.~H.} \&
  \bibinfo{author}{{Miralda-Escud{\'e}}, J.}
\newblock \bibinfo{title}{{Self-Consistent Models of the AGN and Black Hole
  Populations: Duty Cycles, Accretion Rates, and the Mean Radiative
  Efficiency}}.
\newblock \emph{\bibinfo{journal}{\apj}} \textbf{\bibinfo{volume}{690}},
  \bibinfo{pages}{20--41} (\bibinfo{year}{2009}).
\newblock \eprint{0710.4488}.

\bibitem{GeoAkilas19}
\bibinfo{author}{{Georgantopoulos}, I.} \& \bibinfo{author}{{Akylas}, A.}
\newblock \bibinfo{title}{{NuSTAR observations of heavily obscured Swift/BAT
  AGNs: Constraints on the Compton-thick AGNs fraction}}.
\newblock \emph{\bibinfo{journal}{\aap}} \textbf{\bibinfo{volume}{621}},
  \bibinfo{pages}{A28} (\bibinfo{year}{2019}).
\newblock \eprint{1809.03747}.

\bibitem{Ana19}
\bibinfo{author}{{Ananna}, T.~T.} \emph{et~al.}
\newblock \bibinfo{title}{{The Accretion History of AGNs. I. Supermassive Black
  Hole Population Synthesis Model}}.
\newblock \emph{\bibinfo{journal}{\apj}} \textbf{\bibinfo{volume}{871}},
  \bibinfo{pages}{240} (\bibinfo{year}{2019}).
\newblock \eprint{1810.02298}.

\bibitem{Kulier15}
\bibinfo{author}{{Kulier}, A.}, \bibinfo{author}{{Ostriker}, J.~P.},
  \bibinfo{author}{{Natarajan}, P.}, \bibinfo{author}{{Lackner}, C.~N.} \&
  \bibinfo{author}{{Cen}, R.}
\newblock \bibinfo{title}{{Understanding Black Hole Mass Assembly via Accretion
  and Mergers at Late Times in Cosmological Simulations}}.
\newblock \emph{\bibinfo{journal}{\apj}} \textbf{\bibinfo{volume}{799}},
  \bibinfo{pages}{178} (\bibinfo{year}{2015}).

\end{thebibliography}


\setcounter{mybib}{45}

\xpatchcmd{\thebibliography}{%
  \usecounter{enumiv}%
}{%
  \usecounter{enumiv}%
  \setcounter{enumiv}{\value{mybib}}%
}{}{}


\begin{addendum}
 \item FS warmly thanks David Weinberg for a number of illuminating inputs and discussions. FS also acknowledges Dalya Baron, Peter Behroozi, Giorgio Calderone, Benjamin Davis, Fabrizio Fiore, Poshak Gandhi, Sebastian Hoenig, Christian Knigge, Cheng Li, Jordi Miralda-Escud\'{e}, Benjamin Moster, Merry Powell, Ranjan Vasudevan, Carolin Villforth, and Guang Yang for many useful discussions. FS acknowledges partial support from a Leverhulme Trust Research Fellowship and the European Union's Horizon 2020 Programme under the AHEAD project (grant agreement n. 654215). VA acknowledges funding from the European Union's Horizon 2020 research and innovation programme under grant agreement No 749348. MB acknowledges partial support from NSF grant AST-1816330. AL is supported by PRIN MIUR 2017 prot. 20173ML3WW\_002 ``Opening the ALMA window on the cosmic evolution of gas, stars and supermassive black holes''. MK acknowledges support from DLR grant 50OR1904.
\end{addendum}

\noindent \textbf{Author contributions statement}\\
\vspace{-1.1cm}

\begin{addendum}
\item[]
\noindent All authors have contributed in different ways to the work presented in this paper. FS performed the full set up of the AGN mocks, analysis of the results, and writing up of the manuscript. VA independently checked all the results on AGN clustering and largely contributed to text editing and to the referee reports. MB was one of the core Authors in the Shankar et al. (2016) paper on the intrinsic black hole scaling relations and massively contributed to the editing of the manuscript. CM devised accurate determinations of the correlation functions in the MultiDark simulation. AL calculated the accreted black hole mass functions following the models presented in Aversa et al. (2015). NM contributed to the AGN accretion models. PJG and LZ contributed to the galaxy mocks, independently tested some of the key results in the paper, and provided comments. JM performed preliminary large-scale bias estimates of AGN at different luminosities in low redshift galaxies. MK made available a number of data sets on galaxy and AGN clustering inclusive of full covariance matrices. RDB performed independent calculations of some of the AGN mocks. FR contributed to the characterization of the scaling relations in Type 1 and 2 AGN and to the editing of the paper. FLF provided key insights in the use of different AGN bolometric corrections. RKS was one of the core Authors in the Shankar et al. (2016) paper on the intrinsic black hole scaling relations and provided support on the theoretical side and editing of the paper.
\end{addendum}

\noindent \textbf{Additional information}\\
\vspace{-1.1cm}

\begin{addendum}
 \item[Competing Interests] The authors declare that they have no
competing financial interests.
 \item[Correspondence] Correspondence and requests for materials
should be addressed to Francesco Shankar~(email: F.Shankar@soton.ac.uk).
\end{addendum}

\begin{figure*}
\begin{center}
    \center{\includegraphics[width=\textwidth]{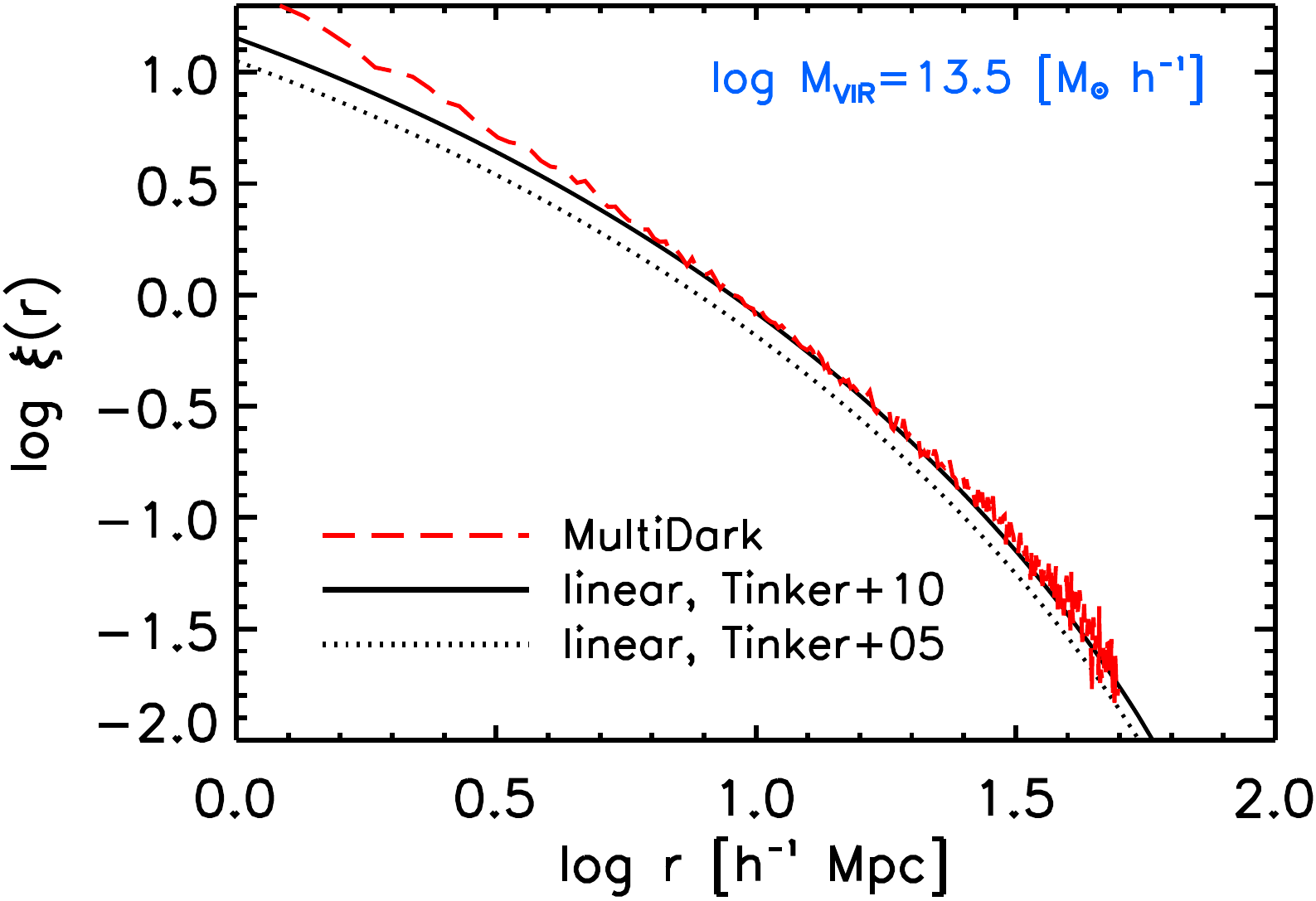}
    \caption{\textbf{Comparing host halo auto-correlation functions.}
    Comparison between the linear matter correlation function multiplied by the square of the halo bias from Tinker et al. (2005; dotted line)\cite{Tinker05} and Tinker et al. (2010; solid line)\cite{Tinker10}, and the autocorrelation function in the MultiDark simulation of all central and satellite haloes with virial mass at infall in the range $13.3<\log M_{\rm vir}/\msune<13.7$. For this comparison we adopt the same cosmological parameters as in the MultiDark simulation.
    \label{fig|BiasHaloMultiDark}}}
\end{center}
\end{figure*}

\begin{figure*}
\begin{center}
    \center{\includegraphics[width=\textwidth]{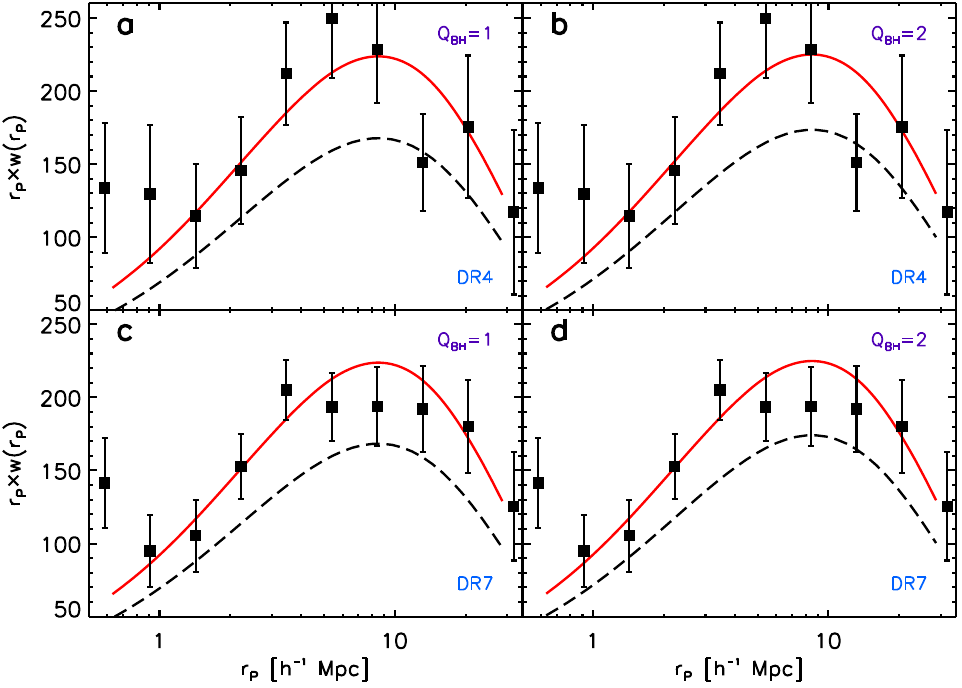}
    \caption{\textbf{Direct comparison with the cross-correlation function of active galaxies.}
    Comparison between the DR4 (top panels) and DR7 (bottom panels) projected AGN-galaxy cross-correlation function derived by Krumpe et al. (filled squares)\cite{Krumpe15}, with the linear matter projected correlation function multiplied by the product of the galaxy and black hole large-scale biases with $Q_{\rm bh}=1$ (left panel) and $Q_{\rm bh}=2$ (right panels). It is clear that the models derived from the intrinsic \mbh-\mstar\ relation (solid red lines) provide a better match to the data (see text for details).
    \label{fig|CCF}}}
\end{center}
\end{figure*}


\begin{table}
\begin{tabular}{|l|l|l|}
\hline
 $\log \mbhe$\,\,\, [$\msune$] & $\log \Phi(\mbhe)\,\,\, [h_{70}^{-3}\, {\rm Mpc^{-3}\, dex^{-1}}]$                                                                         \\ \hline
$6.0$ & $-2.701\pm 0.015$\\
$6.5$ & $-2.730\pm 0.029$\\
$7.0$ & $-2.805\pm 0.045$\\
$7.5$ & $-2.957\pm 0.073$\\
$8.0$ & $-3.218\pm 0.107$\\
$8.5$ & $-3.640\pm 0.148$\\
$9.0$ & $-4.279\pm 0.193$\\
$9.5$ & $-5.233\pm 0.242$\\
$10.$ & $-6.678\pm 0.314$\\
\hline
\end{tabular}
\label{TableBHMF}
\caption{Black hole mass function retrieved from the intrinsic \mbh-\mstar\ relation.}
\end{table}

\end{document}